\documentclass[aip,pof,reprint,superscriptaddress]{revtex4-1}

\usepackage{graphicx}
\usepackage{dcolumn}
\usepackage{bm}

\begin{document}






\title{Cartesian Grid Method for Gas Kinetic Scheme}
\author{Songze Chen}
\affiliation{
The Hong Kong University of Science and Technology \\
Clear Water Bay, Kowloon, Hong Kong, China
}%
\author{Kun Xu}%
\email{makxu@ust.hk}
\affiliation{
The Hong Kong University of Science and Technology \\
Clear Water Bay, Kowloon, Hong Kong, China
}%

\author{Zhihui Li}
\affiliation{
Hypervelocity Aerodynamics Institute, China Aerodynamics Research and Development Center, Mianyang 621000, China
}%

\date{\today}

\begin{abstract}
A Cartesian grid method combined with a simplified gas kinetic scheme is presented for subsonic and supersonic viscous flow simulation on complex geometries.
Under the Cartesian mesh, the computational grid points are classified into four different categories, the fluid point, the solid point, the drop point, and the interpolation point.
The boundaries are represented by a set of direction-oriented boundary points.
A constrained weighted least square method is employed to evaluate the physical quantities at the interpolation points.
Different boundary conditions, including isothermal boundary, adiabatic boundary, and Euler slip boundary, are presented by different interpolation strategies.
We also propose a simplified gas kinetic scheme as the flux solver for both subsonic and supersonic flow computations.
The methodology of constructing a simplified kinetic flux function can be extended to other flow systems.
A few numerical examples are used to validate the Cartesian grid method and the simplified flux function.
The reconstruction scheme   for recovering the boundary conditions of compressible viscous and heat conducting flow with a Cartesian mesh
can provide a smooth distribution of physical quantities at solid boundary, and present an accurate solution for the flow study with complex geometry.

\end{abstract}

\maketitle


\section{Introduction}
Solving flow problem on conformal mesh or unstructured mesh is the most practical strategy in industry applications.
However, the grid generation on complex geometry is time consuming and requires sophisticated technique.
In this sense, the grid generation on complex computational domain is still the bottle neck of practical applications.
Thereby, the Cartesian grid method arises, in which the flow region is discretized by a Cartesian grid regardless of
the shapes of objects inside the flow region.
The obvious advantage of this method over the conventional conformal approach is that the computational mesh is easily generated
when different geometries are considered. Cartesian grid methods free the researchers and engineers from the burdensome grid generation,
but introduce two new problems about boundary treatment.

The first problem with the Cartesian grid is about how to represent the boundary.
It depends on the boundary property. For the multi-fluid and multi-phase flows, where the boundaries or interfaces are deformable,
the volume-of-fluid (VOF) method \cite{Gueyffier1999} and the phase-field approach \cite{Caginalp1989} are the most popular approaches.
The boundary is reconstructed from an auxiliary function, say, a marker function.
Or, the boundary is captured by solving a partial differential equation for the phase field.
These approaches are efficient for boundary deformation and splitting. But obviously, this kind of approaches is less accurate than the Lagrangian representation of the boundary. The latter approach takes the boundary as a sharp interface either explicitly tracking as curves \cite{Udaykumar1999} or as level sets \cite{Chen1997}. Sharp interface is desirable for high Reynolds number flows where the boundary layer plays an important role.


The second problem is how to impose the boundary condition. Following the classification by Mittal and Iaccarino\cite{Mittal2005},
there are two different categories. The first one is the continuous forcing approach;
the second one is the discrete forcing approach.
The continuous forcing approach directly modifies the governing equation by adding a forcing term to take the boundary effect into account\cite{Peskin1982,Goldstein1993}.
An ideal force term is represented by a Dirac delta function. Since the boundary cuts the grid line at arbitrary location, the forcing should be distributed over a band of cells around the boundary point. This approach results in a diffusive boundary. 
The discrete forcing approach imposes the boundary condition on the numerical solution directly at discrete level.
Mittal and Iaccarino \cite{Mittal2005} further subdivided the second category into "Indirect BC Imposition" and "Direct BC Imposition". The former one employs a forcing term which is determined from a priori estimation of flow field at discrete level \cite{Mohd1997}. The external force is explicitly computed in advance or solved by an implicit method to guarantee the no-slip boundary condition \cite{Shu2007,Wu2009}.
However, the artificial forcing procedure diffuses the flow field around the boundary. 
For this reason, the "Direct BC Imposition" approaches retain the boundary as a sharp interface with no spreading.
This can usually be accomplished by modifying the computational stencil near the boundary to directly impose
the boundary condition. This kind of approach is always referred to as Cartesian grid method which is the main focus in this study.

Berger and LeVeque \cite{Berger1989} presented a Cartesian mesh algorithm with adaptive refinement to compute flows around arbitrary geometries by solving the Euler equations.
They treated the intersection between the grid line and the boundaries as a grid point and performed conventional finite difference scheme with amendments of boundary fluxes. However, the small cell instability was observed.
Pember \emph{et al.} \cite{Pember1995} proposed a corrector applied to cells at the boundary to redistribute flow quantities in order to maintain the conservation, therefore, avoid time step restrictions arising from small cells.
In these methods, the amendments or the correctors depend on the specific physical system, thereby, the numerical scheme is difficult for further extension.

Forrer and Jeltsch \cite{Forrer1998} proposed a ghost cell method to apply the boundary condition on Cartesian grid for inviscid compressible Euler equations. The flow quantities at ghost cells which locate inside the boundary are set by their images in the fluid side across the boundary.
As a result, the grid points near the boundary also preserve a complete control volume.
Thus the small cell restriction is removed. Hereafter, this kind of method is referred to as the boundary interpolation method. In recent years, the boundary interpolation method receives more and more attentions.
Udaykumar \emph{et al.} \cite{Udaykumar1999} proposed a finite-difference formulation to track solid-liquid boundaries on a fixed underlying grid. The interface is not of finite thickness but is treated as discontinuity and is explicitly tracked. The imposition of boundary conditions exactly on a sharp interface is performed using simple stencil readjustments in the vicinity of the interface. Ghias \emph{et al.}\cite{Ghias2007} proposed an immersed boundary method for computing viscous, subsonic compressible flows with complex shaped stationary immersed boundaries. The method also employs a ghost cell technique for imposing the boundary conditions on the immersed boundaries. As Gibou \emph{et al.} \cite{Gibou2002} pointed out that the single directional interpolation scheme is poorly behaved for small cells. Therefore, Ghias \emph{et al.} turned to a multidimensional bilinear interpolation and evaluated the flow variables at ghost cell to avoid the small cell problem.

Instead of interpolating variables to the ghost cell which is located inside the solid, some researchers interpolate flow variables at the fluid points, or, the interface points which are very close to the boundary at the fluid side.
Ye \cite{Ye1999} \emph{et al.} and Udaykumar \emph{et al.} \cite{Udaykumar2001} proposed a sharp interface Cartesian grid method for solving the incompressible Navier-Stokes equations with complex moving boundaries. The boundaries cut the mesh, then form some trapezoidal cells. If the trapezoidal cell is very small, then it merges with its neighboring cell, and forms an irregular control volume. The finite volume method can be applied on the merging control volume. Then the small cell problem is circumvented. They proposed an interpolation procedure to construct the flow field near the boundary, so that the boundary condition is satisfied. Tullio \emph{et al.} \cite{Tullio2007} and Palma \emph{et al.} \cite{Palma2006} combines the method for solving the three-dimensional preconditioned Navier-Stokes equations for compressible flows with an immersed boundary approach, to provide a Cartesian-grid method for computing complex flows over a wide range of the Mach number. Moreover, a flexible local grid refinement technique is employed to achieve high resolution near the immersed body and in other high-flow-gradient regions.

Actually, many Cartesian grid method solving the compressible Navier-Stokes equations are based on the interpolation method to handle the boundary conditions.
Peng \emph{et al.} \cite{Peng2008} has compared the forcing method and the interpolation method in the framework of lattice Boltzmann method. They concluded that
the interpolation method perform better than the forcing method. In recent years, more delicate interpolation procedures are employed for boundary treatment.
Seo and Mittal \cite{Seo2011} proposed a new sharp-interface immersed boundary method based approach for the computation of low-Mach number flow-induced sound around complex geometries. Their method applies the boundary condition on the immersed boundary to a high order by combining the ghost cell approach with a weighted least square error method based on a high order approximating polynomial. Duan \emph{et al.}\cite{Duan2010} extended the Cartesian grid method to high order accuracy, and simulated turbulent flow at high Mach numbers.

In this paper, we are going to further develop the boundary interpolation method on Cartesian grid, and combines it with well developed gas kinetic schemes. The main purposes are presented as follows:

1. simplify the gas kinetic schemes, and reduces the computational cost.

2. formulate a constrained weighted least square \cite{Zhang2000} interpolation for different boundary condition.

\section{Simplified gas kinetic scheme}
\label{sec:SGKS}
Gas kinetic schemes (GKS) were developed by Xu \emph{et al} \cite{Xu1995,Xu1998,Xu2001}.
It is based on kinetic theory, using particle distribution function to simulate macroscopic flow motion. Gas kinetic scheme can be considered as a flux solver of a generalized Riemann problem. In this study, the Bhatnagar-Gross-Krook kinetic equation \cite{Bhatnagar1954} is employed.
It takes the following form,
\begin{equation}
    \frac{\partial f}{\partial t}+\mathbf{u}\cdot\frac{\partial f}{\partial \mathbf{x}} =
    \frac{g-f}{\tau},
    \label{eq:BGKModel}
\end{equation}
where $f(\mathbf{x},t,\mathbf{u},\mathbf{\xi})$ represents the particle velocity distribution function. It is a function of location $\mathbf{x}$, particle velocity $\mathbf{u}$, the internal degree of freedom $\mathbf{\xi}$ and time $t$.
The right hand side of Eq.(\ref{eq:BGKModel}) represents the relaxation process. $g$ is the equilibrium state which reads,
\begin{equation}
    g = \mathcal{M}[<f>] \equiv \rho\left(\frac{1}{2\pi RT}\right)^{\frac{K+2}{2}}e^{-\frac{1}{2RT}(\mathbf{u-U})^2},
    \label{eq:MaxwellState}
\end{equation}
where $\rho$ denotes density, $\mathbf{U}$ represents the macroscopic velocity, $T$ denotes temperature, $R$ is gas constant and $K$ is the equivalent internal degree of freedom which is 1 for monatomic gas, and 3 for diatomic gas in two dimensional simulation.
The conservative variables $\mathbf{W}$ can be derived by taking moments of $f$,
\begin{eqnarray}
  \mathbf{W} = \left(\begin{array}{l} \rho \\ \rho \mathbf{U} \\ \rho E \end{array}\right) &=& \int \mathbf{\psi} f d\mathbf{u}d\mathbf{\xi} = <f>\nonumber
\end{eqnarray}
where $\mathbf{\psi} = (1,\mathbf{u}, \frac{1}{2}(\mathbf{u}^2+\mathbf{\xi}^2))^T$, and $E$ denotes the energy per unit mass.
The macroscopic flux of a given distribution function $f$ can be written as,
\begin{eqnarray}
  \mathbf{F} = \int \mathbf{\psi} \mathbf{n}\cdot\mathbf{u} f d\mathbf{u}d\mathbf{\xi} = <\mathbf{n}\cdot\mathbf{u}f>\nonumber,
\end{eqnarray}
where $\mathbf{n}$ denotes a certain direction, say, the normal direction of a surface.

\begin{figure}
    \parbox[b]{0.48\textwidth}{
    \includegraphics[totalheight=6cm, bb = 123 549 331 712, clip =
    true]{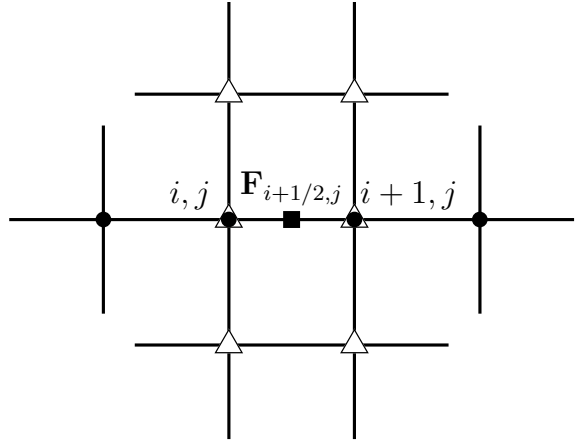}
    }
    \caption{The stencil for numerical flux in x direction.}
    \label{fig:stencilGKS}
\end{figure}

The original GKS is a finite volume method. If the numerical flux $\mathbf{F}$ is known at the cell interfaces of a cell, the macroscopic variables $\mathbf{W}^{n+1}$ for the next time step becomes,
\begin{equation}
\mathbf{W}^{n+1} = \mathbf{W}^{n} - \frac{1}{V}\int_{t^n}^{t^{n+1}}\sum_{k}\mathbf{F}_k S_k dt,
\end{equation}
where $S_k$ denotes the area of the $k$th surface of the cell, and $V$ is the volume of the cell. In this study, we adopt the finite difference formulation. Then, for two dimensional problems,
\begin{equation}
\mathbf{W}^{n+1}_{i,j} = \mathbf{W}^{n}_{i,j} - \int_{t^n}^{t^{n+1}}(\frac{\mathbf{F}_{i+1/2,j}-\mathbf{F}_{i-1/2,j}}{\Delta x}+\frac{\mathbf{F}_{i,j+1/2}-\mathbf{F}_{i,j-1/2}}{\Delta y})dt.
\end{equation}
Figure \ref{fig:stencilGKS} shows the stencil for the numerical flux $\mathbf{F}_{i+1/2,j}$. The solid square represents the flux point, or the cell interface in a finite volume method.
To calculate the numerical flux, the distribution function at the flux point is needed.
The original GKS employs the local analytical solution of the Eq. (\ref{eq:BGKModel}) as the distribution function, which is shown as follows,
\begin{eqnarray}
    f(\mathbf{x}_0,t,\mathbf{u},\mathbf{\xi}) &=& e^{-t/\tau}f(\mathbf{x}_0-\mathbf{u}t,t,\mathbf{u},\mathbf{\xi})  \nonumber \\
        && + \frac{1}{\tau}\int_0^{t}g(\mathbf{x'},t', \mathbf{u},\mathbf{\xi})
        e^{-(t-t')/\tau}dt', \label{eq:localsolution}
\end{eqnarray}
where $\mathbf{x'} = \mathbf{x}_0-\mathbf{u}(t-t')$. This solution includes the complete information from both space and time. The numerical method based on this solution performs well in both continuous and discontinuous flows \cite{Xu2001}. However, the numerical cost is a little higher in comparison with classical Riemann solver-based finite volume method.
And also, a simplified GKS was proposed for low speed and continuous flow simulation in which the Chapman-Enskog expansion is directly employed as the local solution to evaluate the numerical flux at the cell interface \cite{Xu2001},
\begin{eqnarray}
f(t) = g_c - \tau(\mathbf{u}\cdot g_\mathbf{x} + g_t) + g_\mathbf{x}\cdot \mathbf{x} + g_t t. \label{eq:continuousGKS}
\end{eqnarray}
The variables $\mathbf{x}_0, \mathbf{u},\xi$ are omitted hereafter for the sake of simplicity. $g_c$ is the equilibrium state at the cell interface. It is constructed by central difference scheme, namely $g_c = \mathcal{M}[\mathbf{W}_{\mbox{c}}]$. The subscript $\mbox{c}$ denotes the central difference scheme.
This continuous reconstruction is accurate for low speed flow simulation where no shock exists. But, it cannot be used in supersonic flow simulation.

In this study, we replace the leading order term $g_c$ by a non-equilibrium distribution function $f_0$,
\begin{eqnarray}
f(t) = f_0 - \tau(\mathbf{u}\cdot g_\mathbf{x} + g_t) + g_\mathbf{x}\cdot \mathbf{x} + g_t t. \label{eq:discontinuousGKS}
\end{eqnarray}
The first term on the right hand side represents the upwind effect, the second term represents the viscous effect, the third term represents the time evolution up to second order accuracy.
Similar scheme has been proposed \cite{Xu1995}, in which the leading order term is $(1-\alpha)f_0+\alpha g_c$.
Figure \ref{fig:stencilGKS} shows the stencil used in 2D reconstruction. A uniform mesh is assumed.
The non-equilibrium state $f_0$ is
\begin{equation}
f_0 = \left\{\begin{array}{l} g_{l}= \mathcal{M}[\mathbf{W}_{l}],\quad u>0 \\ g_{r}= \mathcal{M}[\mathbf{W}_{r}],\quad u \le 0 \end{array}\right. ,
\end{equation}
where the subscript 'l' and 'r' represent the left and right side of the flux point respectively. $\mathbf{W}_{l}$ and $\mathbf{W}_{r}$ are constructed along the grid line.
For the fluid side, the 3rd order WENO reconstruction is adopted,
\begin{eqnarray}
W_{l} = \frac{w_{-1} W_{-1} + w_{0}W_{0}}{w_{-1}+w_0}, \\
W_{r} = \frac{w_0 W_0 + w_{1}W_{1}}{w_0+w_{1}} ,
\end{eqnarray}
where $W$ represents every single macroscopic quantity of $\mathbf{W}$. The weights are given as follows,
\begin{eqnarray}
w_{-1} &=& \frac{1}{4(s_{i-1}^2+\varepsilon)} , \\
w_0 &=& \frac{3}{4(s_i^2+\varepsilon)} , \\
w_{1} &=& \frac{1}{4(s_{i+1}^2+\varepsilon)} ,
\end{eqnarray}
where $\varepsilon$ is a small positive number to avoid zero denominator, and $s_i = W_{i+1,j}-W_{i,j}$,
\begin{eqnarray}
W_{-1} &=& \frac{3}{2}W_{i,j}-\frac{1}{2}W_{i-1,j} , \\
W_0 &=& \frac{1}{2}W_{i,j}+\frac{1}{2}W_{i+1,j} , \\
W_{1} &=& \frac{3}{2}W_{i+1,j}-\frac{1}{2}W_{i+2,j} .
\end{eqnarray}
For the stencil contains drop point or interpolation point near the boundary, the van-Leer slope limiter is adopted to construct the flow quantities at either side of cell interface.
The spatial derivative in the second term of Eq.(\ref{eq:discontinuousGKS}) can be expressed in the expansion of equilibrium state, which formally reads,
\begin{eqnarray}
    g_{\mathbf{x}} = \mathbf{a}\cdot\mathbf{\psi} \mathcal{M}[\mathbf{W}^{n}],
\end{eqnarray}
where the expansion coefficient $\mathbf{a}$ reads,
\begin{eqnarray}
    \mathbf{a}\cdot\mathbf{\psi} = a_{0,d}+\mathbf{a}_{1,d}\cdot \mathbf{u}+a_{2,d}\frac{\mathbf{u}^2+\mathbf{\xi}^2}{2}, \ \ d = 1,...,D.
\end{eqnarray}
where $\mathbf{a}_{1,d} = (a_{1,d1}, ..., a_{1,dD})$, and $D$ denotes the dimension. For example, for two dimensional problems, $D = 2$.
Then take the conservative moments of the spatial derivatives of equilibrium state, we have,
\begin{eqnarray}
    \mathbf{a}\cdot<\mathbf{\psi}\mathcal{M}[\mathbf{W}^{n}]> = \mathbf{W}_{\mathbf{x}}, \label{eq:derivativeMatrix}
\end{eqnarray}
where $\mathbf{W}^{n}$ denotes the conservative variables at the beginning of the time step. So, the corresponding conservative variables are $\mathbf{W}^n = <f_0>$ for the Eq. (\ref{eq:discontinuousGKS}).
Then $\mathbf{W}_{\mathbf{x}}$ is derived by a central difference scheme. Take two dimensional problems as an example,
\begin{eqnarray}
\mathbf{W}_x &=& \frac{\mathbf{W}_{i+1,j}-\mathbf{W}_{i,j}}{\Delta x} \\
\mathbf{W}_y &=& \frac{1}{4\Delta y}(\mathbf{W}_{i,j+1}+\mathbf{W}_{i+1,j+1}-\mathbf{W}_{i,j-1}-\mathbf{W}_{i+1,j-1})
\end{eqnarray}
Solving the Eq. (\ref{eq:derivativeMatrix}), we get the expression of $g_{\mathbf{x}}$.

The temporal derivative is estimated by the direct difference of the equilibrium state on the time interval.
\begin{eqnarray}
    g_{t} &=& \frac{1}{\Delta t}(\mathcal{M}[\mathbf{W}^{n+1}] - \mathcal{M}[\mathbf{W}^{n}]), \nonumber \\
\end{eqnarray}
where $\mathbf{W}^{n+1}$ denotes the conservative variables at the end of the time step.
It can be derived by applying the conservation constraint, $<\tau(g_{t} + \mathbf{u}\cdot g_{\mathbf{x}})> = 0$ at a cell interface, that is,
\begin{eqnarray}
    \mathbf{W}^{n+1} = \mathbf{W}^{n} - \Delta t <\mathbf{u}\cdot g_{\mathbf{x}}>.
\end{eqnarray}
The time dependent distribution function at flux point is,
\begin{eqnarray}
    f(t) &=&
     f_0 - \tau(\mathbf{u}\cdot g_{x}+\frac{\mathcal{M}[\mathbf{W}^{n+1}]-\mathcal{M}[\mathbf{W}^{n}]}{\Delta t}) + t\frac{\mathcal{M}[\mathbf{W}^{n+1}]-\mathcal{M}[\mathbf{W}^{n}]}{\Delta t}\nonumber \label{eq:discreteLocalSolution}
\end{eqnarray}
Therefore, the numerical flux reads,
\begin{eqnarray}
    \int_{t^n}^{t^{n+1}}\mathbf{F}dt &=& \mathbf{n}\cdot\int_{t^n}^{t^{n+1}}\int \psi \mathbf{u} f(t)d\mathbf{u}d\mathbf{\xi} dt \nonumber\\
     &=& \mathbf{n}\cdot(\Delta t<\mathbf{u}f_0> - \tau \Delta t <\mathbf{u}\mathbf{u}\cdot g_{\mathbf{x}}> \\
     && + (\frac{\Delta t}{2}- \tau) <\mathbf{u}(\mathcal{M}[\mathbf{W}^{n+1}]-\mathcal{M}[\mathbf{W}^{n}])>),\nonumber \label{eq:localsolution}
\end{eqnarray}
Up to this point, the simplified gas kinetic scheme is completed. Similar scheme has been studied in \cite{XuJ1995}.

\section{Cartesian grid method}
\begin{figure}
\centering
    \includegraphics[totalheight=6cm, bb = 117 516 447 718, clip =
    true]{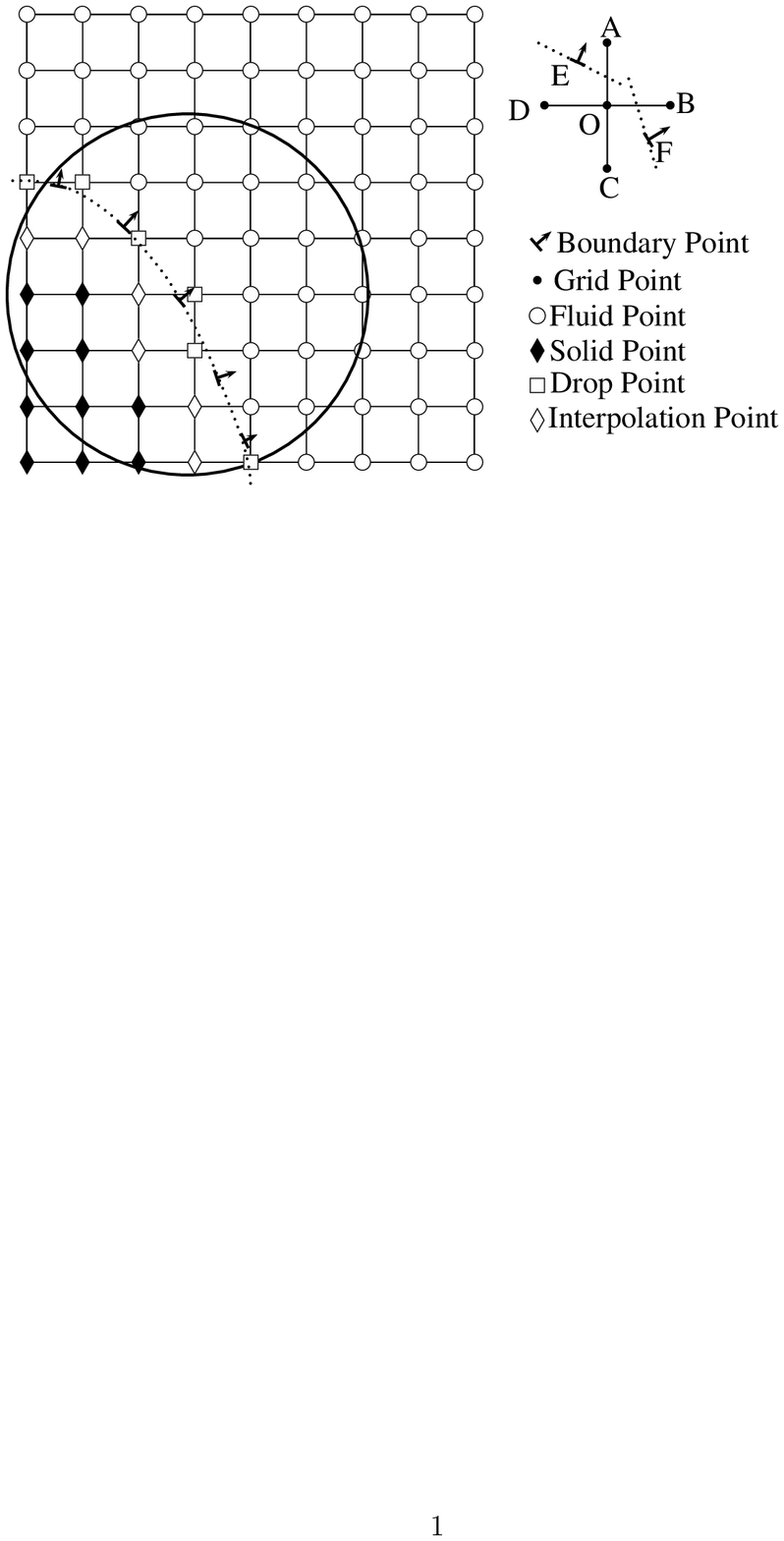}
    \caption{The stencil for the boundary interpolation.}
    \label{fig:stencilBoundary}
\end{figure}

Figure \ref{fig:stencilBoundary} illustrates the stencil for the boundary interpolation. The circles represent the fluid points; the squares represent the drop points which are excluded from the flux stencil
and the interpolation stencil; the diamonds represent the solid points, while the empty diamonds are the interpolation points or ghost points;
and the boundary points are denoted by  points with their normal directions.
The right top part of figure \ref{fig:stencilBoundary} illustrates how to classify a grid point.
The grid points (O,D,C) are solid points, since they locate at the back of the neighboring boundary points (E and F).
The grid points A and B are fluid points, which locate at the front of boundary points (E or F).
If any grid point is too close to the boundary point, then, this point is marked as a drop point. When a solid point is involved in a flux stencil, then it becomes an interpolation point.
After defining different kind of grid points,
all the flux stencils are composed of interpolation points and fluid points.
Meanwhile, the small cells formed by the cutting process are totally avoided, and no stability problem is observed.

The boundary condition is satisfied by assigning the corresponding values at the interpolation points or ghost points.
The solid points which are involved in a flux stencil are employed as interpolation points. It means that an interpolation point can preserve different values when it is in different flux stencil. To satisfy the boundary condition, the fluid variables at the interpolation point should be assigned by appropriate interpolation.
As shown in figure \ref{fig:stencilBoundary}, we use the fluid points inside the big circle to reconstruct the fluid variables at interpolation points to enforce the corresponding boundary conditions.

\subsection{Constrained weighted least square interpolation}
Seo\cite{Seo2011} proposed a weighted least square interpolation for low Mach number acoustic simulation.
For low Mach number flow, the boundary value can be predetermined, since the density is constant, the velocity is given at boundary, and the pressure is totally determined by a Poisson equation.
This is not the case for the compressible viscous flow simulation.
The boundary variables are unknown before the interpolation. As a result, the weighted least square method cannot be used in compressible flow simulation directly.
The boundary variables should be predetermined from the neighboring fluid properties.
For example, Palma \emph{et al.} \cite{Palma2006} assumed the constant pressure across the first layer of fluid point near the boundary,
then use the derived boundary variables to construct the fluid variables at the interpolation point. In this study, we employ a constrained weighted least square method \cite{Zhang2000} to construct the fluid variables at the interpolation point. This interpolation process interpolates an intermediate flow state first, which is subsequently modified in order to satisfy the boundary condition. Then,
the distribution of flow variables around the boundary is determined.
Assume that the fluid variable around the boundary can be expressed as a polynomial,
\begin{eqnarray}
P(x,y) = \sum_{p+q \le m} c_{p,q}x^p y^q, p \ge 0, q \ge 0,
\end{eqnarray}
 where $m$ is the order of polynomial. Given the values $\{v_i\}$ are known at a set of locations $\{(x_i,y_i)|i = 1,...,N, N \ge \mbox{dim}(\mathbf{X})\}$.
And they should be approximately satisfied in the sense of minimizing a weighted mean square error function, which reads,
\begin{eqnarray}
\mbox{minimize}\ \ \sum_{i = 1}^{N} (w_i(\mathbf{A}_i\mathbf{X} - v_i))^2. \label{eq:leastSquareMini}
\end{eqnarray}
where $w_i$ is the weight which is written in an exponential form,
\begin{equation}
w_i = \exp(-\frac{\beta^2((x_i-x_b)^2+(y_i-y_b)^2)}{(\max(\Delta x,\Delta y))^2}),
\end{equation}
where $\beta$ is an adjustable parameter. And $v_i$ denotes the every single variable of $\mathbf{W}_i$.
\begin{eqnarray}
\mathbf{X} &=& (c_{0,0},\ ...,\ c_{m,0},\ c_{0,1},\ ...,\ c_{m-1,1},\ ...,\ c_{0,m})^{T}, \\
\mathbf{A}_i &=& (1,\ ...,\ x_i^m y_i^0,\ x_i^0 y_i^1,\ ...,\ x_i^{m-1} y_i^1,\ x_i^0 y_i^m).
\end{eqnarray}
At first, the unconstrained problem described above is solved. Let $\mathbf{v} = (v_1,v_2,...,v_N)$,
$$
\mathbf{w} = \left(\begin{array}{c@{\ }c@{\ }c@{\ }c}
w_0 & 0 & ... & 0 \\
0 & w_1 & ... & 0 \\
... & ... & ... & ... \\
0 & 0 & ... & w_N
\end{array}\right),\  \mbox{and}\ \mathbf{A} = \left(\begin{array}{c}\mathbf{A}_1 \\ \mathbf{A}_2 \\ ... \\ \mathbf{A}_N\end{array}\right).
$$
Then the solution is expressed as follows,
\begin{eqnarray}
\mathbf{X} &=& \mathbf{C}^{-1}\mathbf{A}^{T}\mathbf{w}^2 \mathbf{v},
\end{eqnarray}
where $\mathbf{C} = \mathbf{A}^T \mathbf{w}^2 \mathbf{A}$.
And the value at boundary point $(x_b,y_b)$ is evaluated by the value of the unconstrained polynomial $P^*(x_b,y_b)$.

Using this interpolation process, the intermediate boundary variable $\mathbf{W}^*_b$ is calculated. Then, to determine the pressure on the boundary, a second intermediate boundary density and pressure are calculated by
\begin{eqnarray}
\left.\begin{array}{lcc}
\rho^{**}_b &=& \rho^*_b (1+\frac{1}{K+2}\mbox{Ma}_{\mathbf{n}}^2)^{\frac{K+2}{2}} \\
p^{**}_b &=& p^*_b (1+\frac{1}{K+2}\mbox{Ma}_{\mathbf{n}}^2)^{\frac{K+4}{2}}
\end{array}\right\} &\ \ & \mbox{for leeward surface where} (\mathbf{U}^*_b-\mathbf{U}_{wall})\cdot\mathbf{n} > 0 , \\
\left.\begin{array}{lcc}
\rho^{**}_b &=& \rho^*_b \\
p^{**}_b &=& p^*_b
\end{array}\right\} &\ \ & \mbox{for the other situation,}
\end{eqnarray}
where $\mathbf{n}$ denotes the normal direction of a boundary surface, and $\mbox{Ma}_{\mathbf{n}}$ denotes the Mach number defined by the ratio of the orthogonal relative velocity to the boundary and the local speed of sound.
The density and pressure assigned at the leeward surface are extremely important for supersonic flow, since the extrapolation from the downstream to the leeward surface is physically invalid at supersonic speed.
Therefore, the stagnation point condition for one dimensional isentropic flow is adopted to construct the upstream flow condition at the leeward surface.
Obviously, the intermediate boundary velocity is not
the no-slip boundary condition or no-penetration boundary condition. Then, $\mathbf{W}^{**}_b$ is transformed to $\mathbf{W}_b$ to satisfy the boundary condition.
To implement the isothermal no-slip boundary condition, we also assume that the pressure derived from $\mathbf{W}_b$ is identical to the pressure derived from $\mathbf{W}^{**}_b$.
And the velocity and the temperature are replaced by the boundary velocity and temperature respectively. That is,
\begin{eqnarray}
p_b & = & p_b^{**}  , \\
\mathbf{U}_b & = & \mathbf{U}_{wall} ,\\
\rho_b &=& p^{**}_b/(RT_{wall}) ,\\
\rho_b E_b &=& \frac{1}{2}\rho_b \mathbf{U}_b^2 + \frac{K+2}{2}p^{**}_b .
\end{eqnarray}
The Euler slip boundary condition (reflection boundary condition) presents,
\begin{eqnarray}
p_b &=& p^{**}_b , \\
\mathbf{U}_b & = & (\mathbf{n}\cdot\mathbf{U}_{wall})\mathbf{n} + \mathbf{U}^{*}_b-(\mathbf{n}\cdot\mathbf{U}^{*}_b)\mathbf{n} , \\
\rho_b &=& \rho^{**}_b , \\
\rho_b E_b &=& \frac{1}{2}\rho_b \mathbf{U}_b^2 + \frac{K+2}{2}p^{**}_b .
\end{eqnarray}
The no-slip adiabatic boundary condition gives,
\begin{eqnarray}
p_b &=& p^{**}_b , \\
\mathbf{U}_b & = & \mathbf{U}_{wall} , \\
\rho_b &=& \rho^{**} , \\
\rho_b E_b &=& \frac{1}{2}\rho_b \mathbf{U}_b^2 + \frac{K+2}{2}p^{**}_b .
\end{eqnarray}
In this study, the solid boundary is stationary, thus $\mathbf{U}_{wall} = 0$.
Then, the boundary variables $\mathbf{W}_b$ at several boundary points $\{(x_{b,j},y_{b,j})|j = 1,...,M, M <  \mbox{dim}(\mathbf{X})\}$ must satisfy the constraint,
\begin{eqnarray}
\mbox{Constraint}:\ \  \mathbf{A}_b\mathbf{X} = \mathbf{v}_b,
\end{eqnarray}
where $\mathbf{v}_b = (v_{b,1}, v_{b,2}, ..., v_{b,M})$, $v_{b,j}$ denotes the every single component of $\mathbf{W}_{b,j}$, and
\begin{eqnarray}
\mathbf{A}_b &=& \left(\begin{array}{c@{\ ...\ }cc@{\ ...\ }c@{\ ...\ }c}
1 & x_{b,1}^m y_{b,1}^0 & x_{b,1}^0 y_{b,1}^1 & x_{b,1}^{m-1} y_{b,1}^1 & x_{b,1}^0 y_{b,1}^m \\
1 & x_{b,2}^m y_{b,2}^0 & x_{b,2}^0 y_{b,2}^1 & x_{b,2}^{m-1} y_{b,2}^1 & x_{b,2}^0 y_{b,2}^m \\
... & ... & ... & ... & ... \\
1 & x_{b,M}^m y_{b,M}^0 & x_{b,M}^0 y_{b,M}^1 & x_{b,M}^{m-1} y_{b,M}^1 & x_{b,M}^0 y_{b,M}^m
\end{array}\right) .
\end{eqnarray}
Combining with Eq.(\ref{eq:leastSquareMini}), the constrained solution is
\begin{eqnarray}
\mathbf{X} &=& \mathbf{C}^{-1}(I - \mathbf{A}_b^{T} \mathbf{C}_b^{-1}\mathbf{A}_b\mathbf{C}^{-1})\mathbf{A}^{T}\mathbf{w}^2 \mathbf{v}
    + \mathbf{C}^{-1} \mathbf{A}_b^{T} \mathbf{C}_b^{-1}\mathbf{v}_b, \\
&&\mbox{with}\ \  \mathbf{C}_b =  \mathbf{A}_b \mathbf{C}^{-1} \mathbf{A}_b^{T}.
\end{eqnarray}
So we fully determine the polynomial $P(x,y)$ at the fluid side.
If only with the consideration of one boundary point, then the current interpolation reproduces the least square method employed by Seo \cite{Seo2011}.
Furthermore, the unconstrained solution can be reused for calculating the constrained solution (for instance, $\mathbf{C}^{-1}$).
Therefore, when the boundary variables cannot even be predetermined before the interpolation,
the current interpolation method is identically efficient.
In this study we only consider the linear polynomial interpolation.
According to different boundary conditions, the derivatives at the interpolation points (or ghost points) on the solid side are determined
by applying symmetric boundary condition. For the isothermal no-slip boundary,
\begin{eqnarray}
\frac{\partial \rho}{\partial \mathbf{n}}|_{\mbox{fluid}} = \frac{\partial \rho}{\partial \mathbf{n}}|_{\mbox{solid}}&,\ \ &
    \frac{\partial \rho}{\partial \mathbf{t}}|_{\mbox{fluid}} = \frac{\partial \rho}{\partial \mathbf{t}}|_{\mbox{solid}}, \\
\frac{\partial \mathbf{U}_{\mathbf{n}}}{\partial \mathbf{n}}|_{\mbox{fluid}} = \frac{\partial \mathbf{U}_{\mathbf{n}}}{\partial \mathbf{n}}|_{\mbox{solid}}&,\ \ &
    \frac{\partial \mathbf{U}_{\mathbf{n}}}{\partial \mathbf{t}}|_{\mbox{fluid}} = \frac{\partial \mathbf{U}_{\mathbf{n}}}{\partial \mathbf{t}}|_{\mbox{solid}}, \\
\frac{\partial \mathbf{U}_{\mathbf{t}}}{\partial \mathbf{n}}|_{\mbox{fluid}} = \frac{\partial \mathbf{U}_{\mathbf{t}}}{\partial \mathbf{n}}|_{\mbox{solid}}&,\ \ &
    \frac{\partial \mathbf{U}_{\mathbf{t}}}{\partial \mathbf{t}}|_{\mbox{fluid}} = \frac{\partial \mathbf{U}_{\mathbf{t}}}{\partial \mathbf{t}}|_{\mbox{solid}}, \\
\frac{\partial (\rho E)}{\partial \mathbf{n}}|_{\mbox{fluid}} = \frac{\partial (\rho E)}{\partial \mathbf{n}}|_{\mbox{solid}}&,\ \ &
    \frac{\partial (\rho E)}{\partial \mathbf{t}}|_{\mbox{fluid}} = \frac{\partial (\rho E)}{\partial \mathbf{t}}|_{\mbox{solid}};
\end{eqnarray}
for the Euler slip boundary,
\begin{eqnarray}
\frac{\partial \rho}{\partial \mathbf{n}}|_{\mbox{fluid}} = -\frac{\partial \rho}{\partial \mathbf{n}}|_{\mbox{solid}}&,\ \ &
    \frac{\partial \rho}{\partial \mathbf{t}}|_{\mbox{fluid}} = \frac{\partial \rho}{\partial \mathbf{t}}|_{\mbox{solid}}, \\
\frac{\partial \mathbf{U}_{\mathbf{n}}}{\partial \mathbf{n}}|_{\mbox{fluid}} = \frac{\partial \mathbf{U}_{\mathbf{n}}}{\partial \mathbf{n}}|_{\mbox{solid}}&,\ \ &
    \frac{\partial \mathbf{U}_{\mathbf{n}}}{\partial \mathbf{t}}|_{\mbox{fluid}} = -\frac{\partial \mathbf{U}_{\mathbf{n}}}{\partial \mathbf{t}}|_{\mbox{solid}}, \\
\frac{\partial \mathbf{U}_{\mathbf{t}}}{\partial \mathbf{n}}|_{\mbox{fluid}} = -\frac{\partial \mathbf{U}_{\mathbf{t}}}{\partial \mathbf{n}}|_{\mbox{solid}}&,\ \ &
    \frac{\partial \mathbf{U}_{\mathbf{t}}}{\partial \mathbf{t}}|_{\mbox{fluid}} = \frac{\partial \mathbf{U}_{\mathbf{t}}}{\partial \mathbf{t}}|_{\mbox{solid}}, \\
\frac{\partial (\rho E)}{\partial \mathbf{n}}|_{\mbox{fluid}} = -\frac{\partial (\rho E)}{\partial \mathbf{n}}|_{\mbox{solid}}&,\ \ &
    \frac{\partial (\rho E)}{\partial \mathbf{t}}|_{\mbox{fluid}} = \frac{\partial (\rho E)}{\partial \mathbf{t}}|_{\mbox{solid}};
\end{eqnarray}
for the adiabatic no-slip boundary,
\begin{eqnarray}
\frac{\partial \rho}{\partial \mathbf{n}}|_{\mbox{fluid}} = -\frac{\partial \rho}{\partial \mathbf{n}}|_{\mbox{solid}}&,\ \ &
    \frac{\partial \rho}{\partial \mathbf{t}}|_{\mbox{fluid}} = -\frac{\partial \rho}{\partial \mathbf{t}}|_{\mbox{solid}}, \\
\frac{\partial \mathbf{U}_{\mathbf{n}}}{\partial \mathbf{n}}|_{\mbox{fluid}} = \frac{\partial \mathbf{U}_{\mathbf{n}}}{\partial \mathbf{n}}|_{\mbox{solid}}&,\ \ &
    \frac{\partial \mathbf{U}_{\mathbf{n}}}{\partial \mathbf{t}}|_{\mbox{fluid}} = \frac{\partial \mathbf{U}_{\mathbf{n}}}{\partial \mathbf{t}}|_{\mbox{solid}}, \\
\frac{\partial \mathbf{U}_{\mathbf{t}}}{\partial \mathbf{n}}|_{\mbox{fluid}} = \frac{\partial \mathbf{U}_{\mathbf{t}}}{\partial \mathbf{n}}|_{\mbox{solid}}&,\ \ &
    \frac{\partial \mathbf{U}_{\mathbf{t}}}{\partial \mathbf{t}}|_{\mbox{fluid}} = \frac{\partial \mathbf{U}_{\mathbf{t}}}{\partial \mathbf{t}}|_{\mbox{solid}}, \\
\frac{\partial (\rho E)}{\partial \mathbf{n}}|_{\mbox{fluid}} = -\frac{\partial (\rho E)}{\partial \mathbf{n}}|_{\mbox{solid}}&,\ \ &
    \frac{\partial (\rho E)}{\partial \mathbf{t}}|_{\mbox{fluid}} = -\frac{\partial (\rho E)}{\partial \mathbf{t}}|_{\mbox{solid}}£¬
\end{eqnarray}
where $\frac{\partial}{\partial \mathbf{n}}$ and $\frac{\partial}{\partial \mathbf{t}}$ denote the partial derivatives along the normal direction
and the tangential direction of boundary surface respectively. Up to now, the linear interpolation polynomial at the solid side is totally determined, so is the variable at the interpolation point.
This interpolation procedure is adopted for both low speed and supersonic flow simulations.

\section{Numerical results}

This section includes numerical examples from the subsonic incompressible limit to the hypersonic compressible viscous flow computations.

\subsection{Subsonic}
The isothermal no-slip boundary condition is adopted for all the subsonic flow simulations.
\subsubsection{Lid-driven cavity flow}
The lid-driven cavity flow is a classical benchmark problem.
With the Cartesian grid method, it  is not easy to simulate internal flows,
since the method theoretically cannot fully guarantee the mass conservation in the computational domain.
Even though, it is still valuable to try the cavity case to estimate the conservative property of the Cartesian grid method.

\begin{figure}
    \parbox[b]{0.48\textwidth}{
    \includegraphics[totalheight=6cm, bb = 90 35 690 575, clip =
    true]{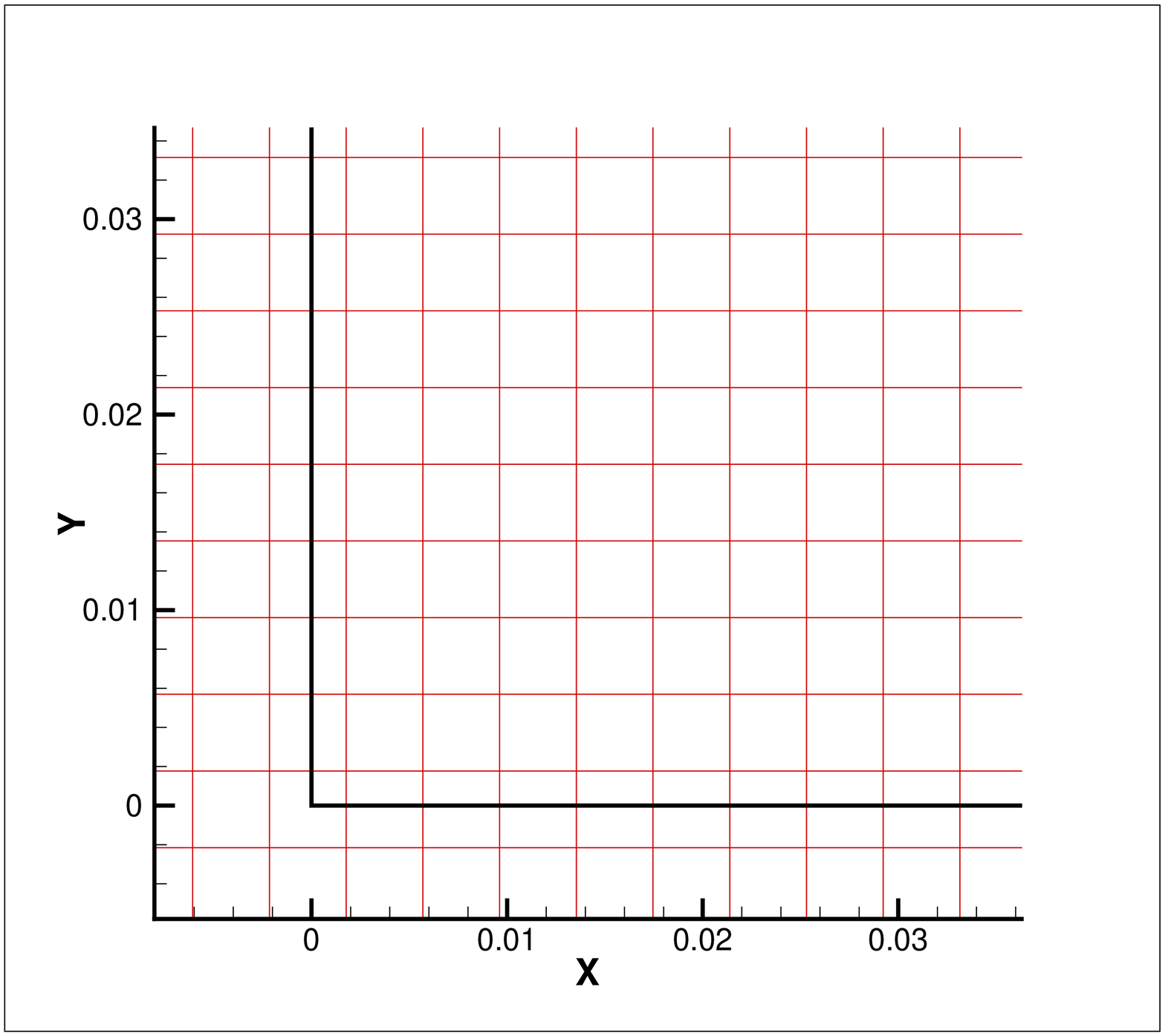}
    }
    \parbox[b]{0.48\textwidth}{
    \includegraphics[totalheight=6cm, bb = 90 35 690 575, clip =
    true]{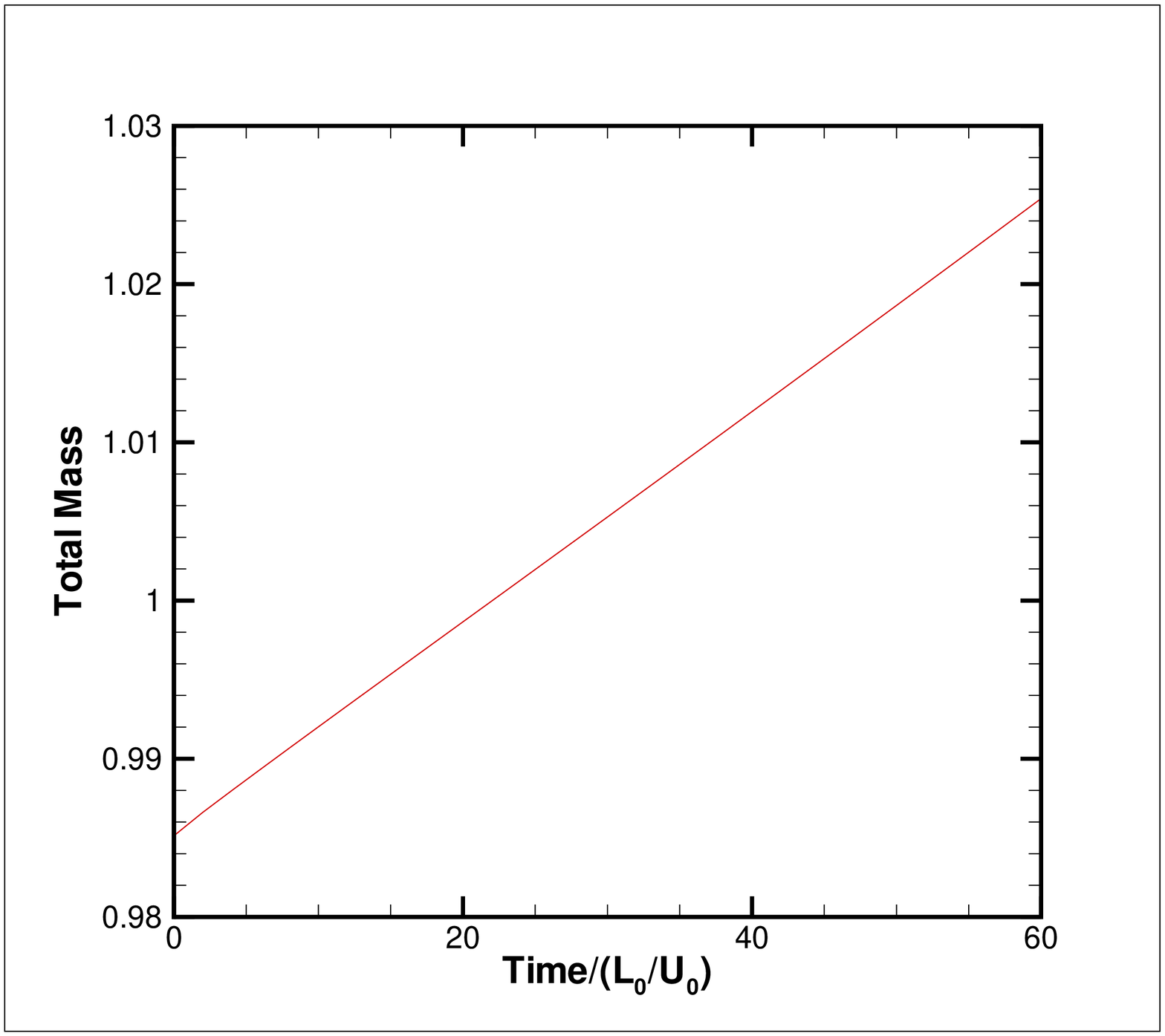}
    }
    \caption{The mesh at left bottom corner (left figure). The time series of the total mass of all the fluid points for the lid-driven cavity flow at $\mbox{Re} = 1000$ (right figure).}
    \label{fig:cavityMass} 
\end{figure}

\begin{figure}
    \parbox[b]{0.48\textwidth}{
    \includegraphics[totalheight=6cm, bb = 90 35 690 575, clip =
    true]{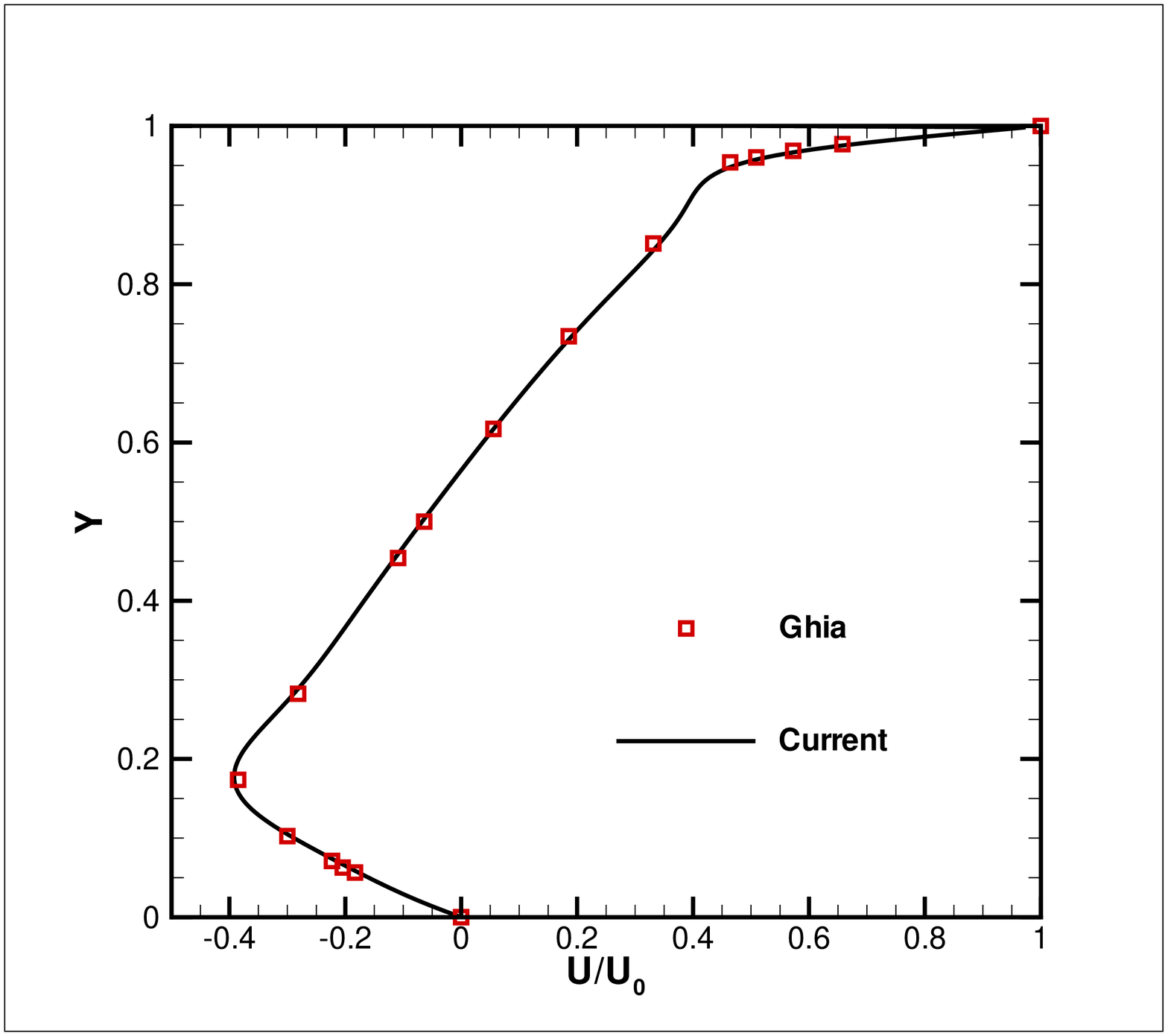}
    }
    \hfill
    \parbox[b]{0.48\textwidth}{
    \includegraphics[totalheight=6cm, bb = 90 35 690 575, clip =
    true]{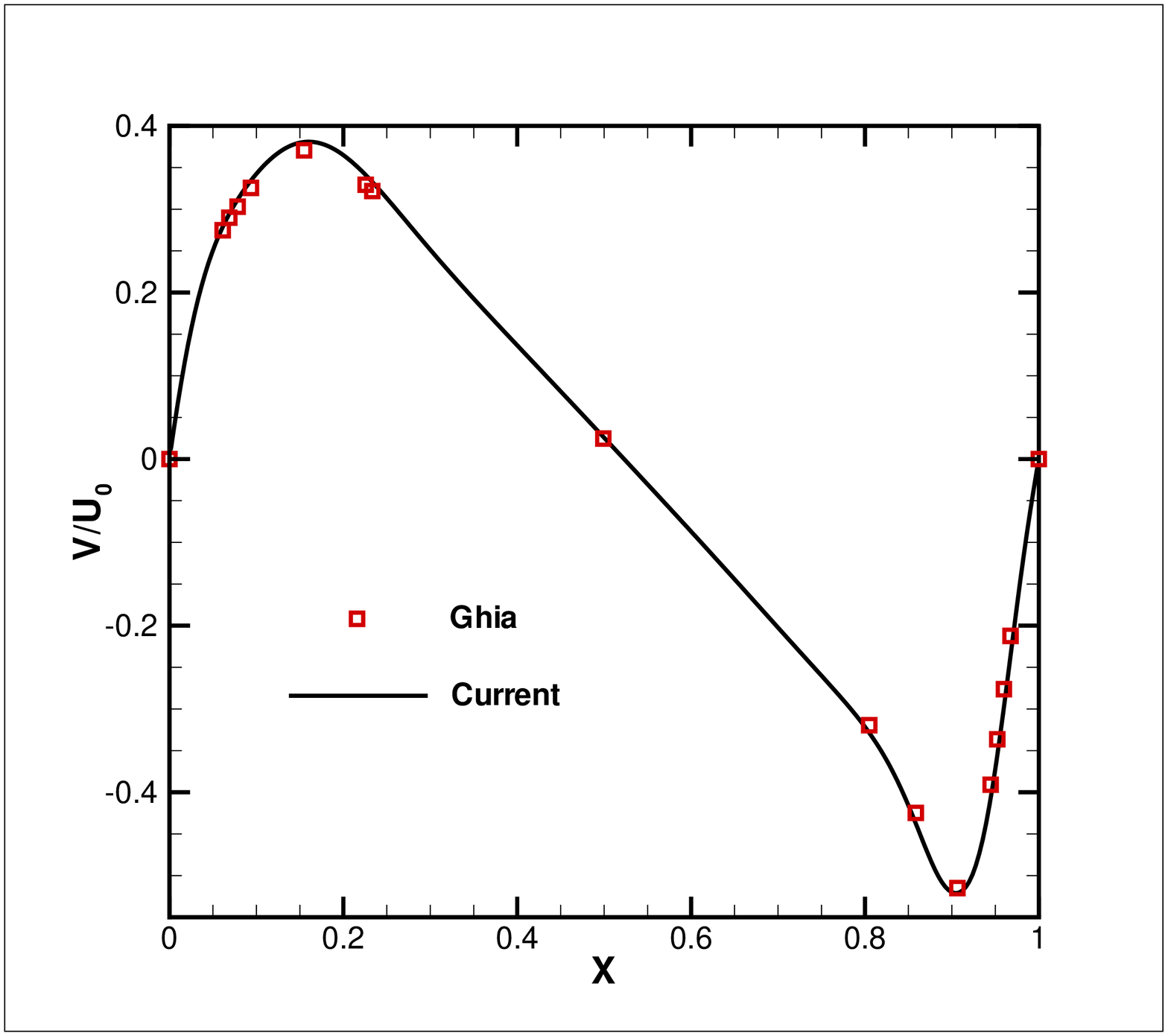}
    }
    \caption{The velocity U profile along the vertical central line and the velocity V profile along the horizontal central line for the lid-driven cavity flow at $\mbox{Re} = 1000$. The grid is $261 \times 261$ covering the range of $[-0.01,1.01]\times[-0.01,1.01]$ }
    \label{fig:cavityUV} 
\end{figure}

The computational region is defined on a domain of $[-0.01,1.01]\times[-0.01,1.01]$,
where two vertical walls locate at $x=0$ and $x=1$, two horizontal walls locate at $y = 0$ and $y = 1$.
The length of the cavity is $L_0 = 1$. The upper wall is moving from left to right with a constant velocity.
The isothermal boundary condition is adopted. The boundary values are given as follows,
\begin{eqnarray}
    \mathbf{U} = (U_{0},0),\ \ T = 1 &, & \quad 0<x<1,y = 1 , \\
    \mathbf{U} = (0,0),\ \ T = 1 &, & \quad \mbox{the other boundaries}.
\end{eqnarray}
The number of grid points in each direction is $261$. Since a margin adjacent to the solid wall is preserved, the internal area of cavity is approximately discretized by a $256\times 256$ mesh (see the left bottom of the cavity in figure \ref{fig:cavityMass}). For the later simulations, the vertical and horizontal boundaries are cut by the grid lines similarly.
The density is 1, and the temperature is 1 at the beginning.
The figure \ref{fig:cavityMass} shows the total mass inside the cavity.
The total mass should be a constant for a steady flow. However, the constant slope of the profile indicates that the total mass increases with time increasing.
The gain or loss of the total mass depends on the relative position of the grid point to the boundary point and the specific boundary treatment.
The mass changing rate is about 0.1\% per unit time for this test case.
The numerical results are compared with the Ghia's reference value. The velocity profiles at time $t=30L_0/U_0$, when the total mass is not changing too much, are plotted in the figure \ref{fig:cavityUV}.

Although the boundary condition is not fully conservative, the velocity profiles are still quit good.
The viscous effect can be simulated correctly in this case.

\subsubsection{Flow past a circular cylinder}
\begin{figure}
    \parbox[b]{0.48\textwidth}{
    \includegraphics[totalheight=6cm, bb = 90 35 690 575, clip =
    true]{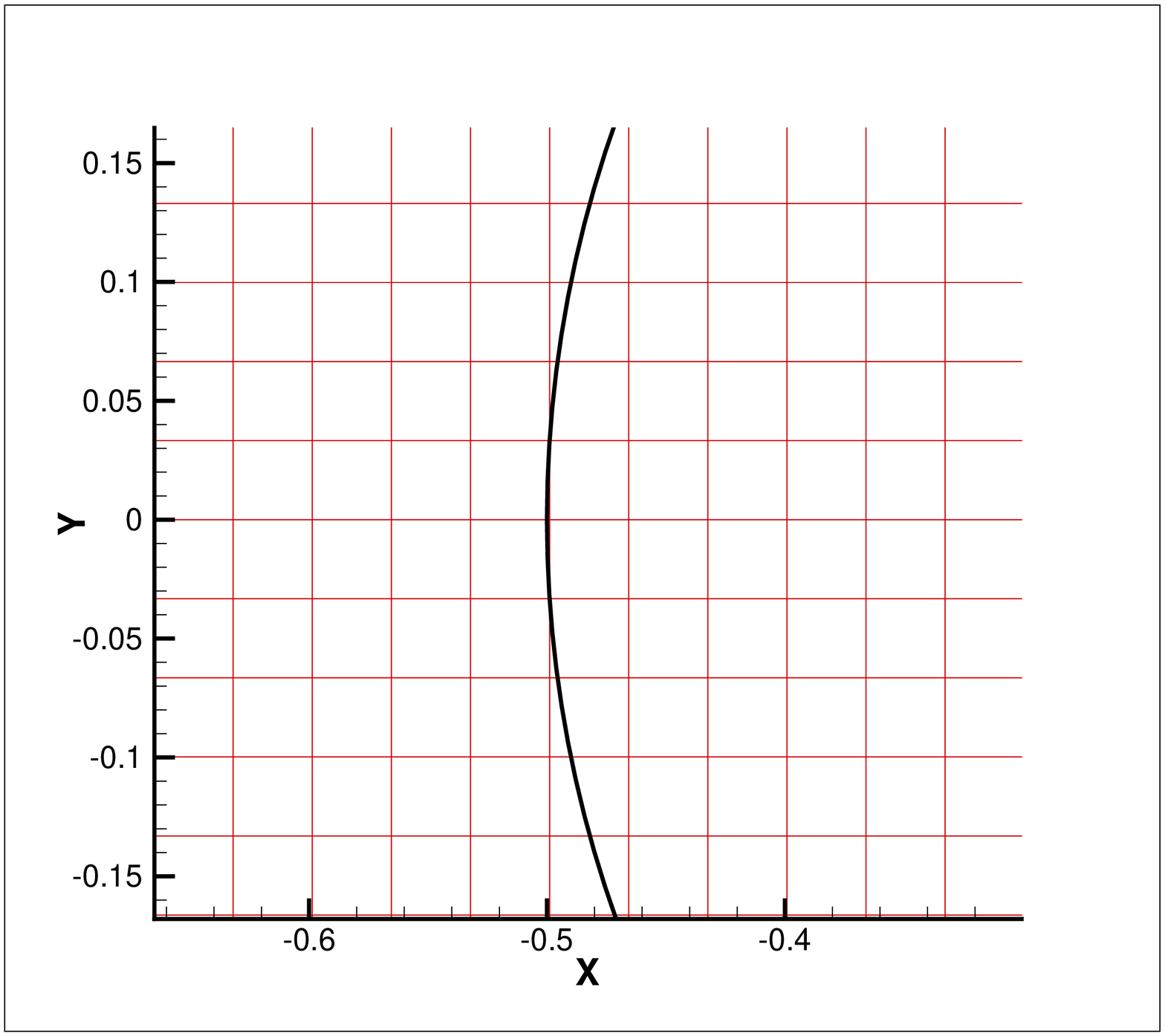}
    }
    \parbox[b]{0.48\textwidth}{
    \includegraphics[totalheight=6cm, bb = 90 35 690 575, clip =
    true]{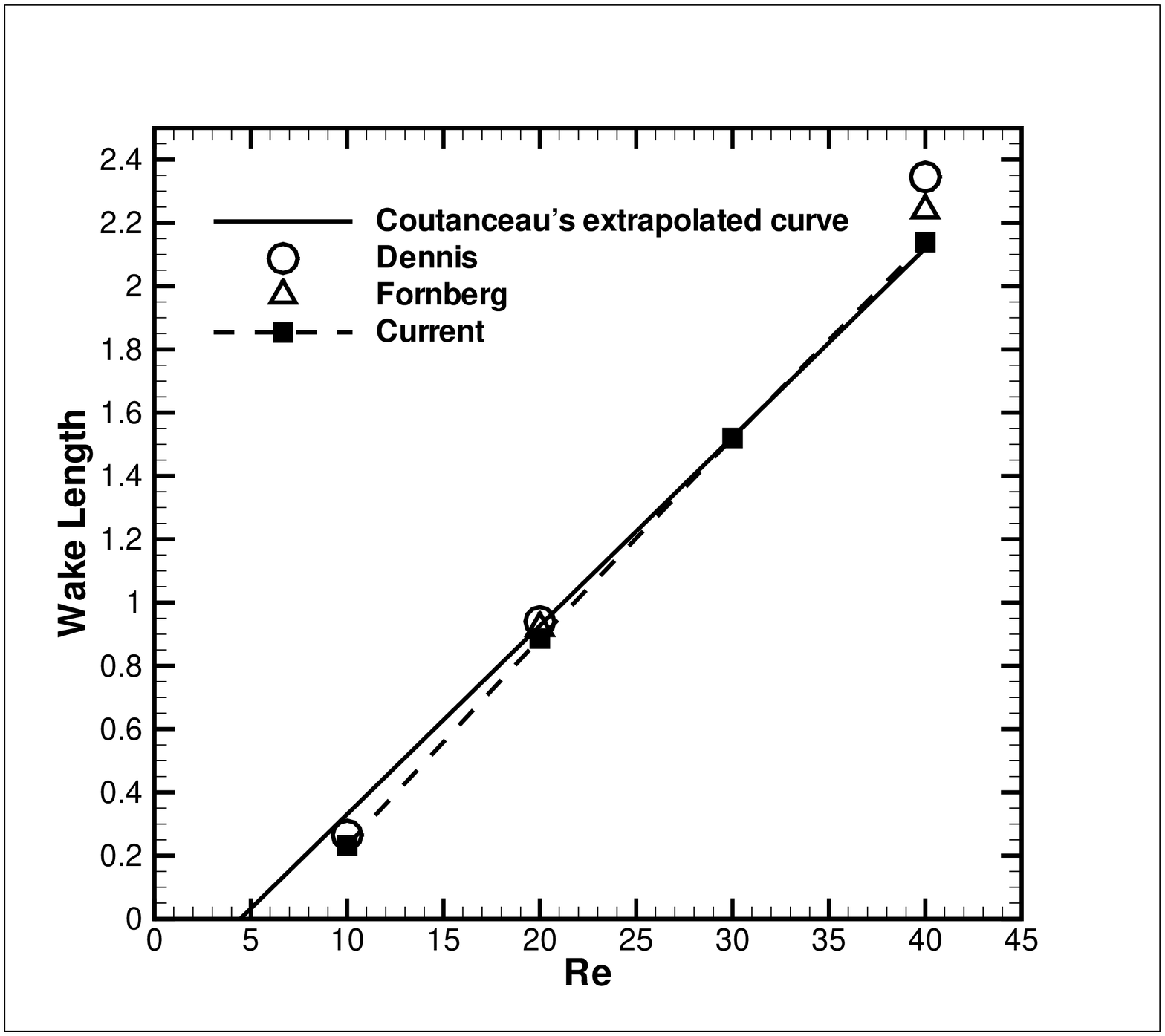}
    }
    \caption{The mesh around the cylinder (left figure). The wake length vs. Reynolds number for the flow past a circular cylinder (right figure).}
    \label{fig:circleWakeLength} 
\end{figure}

\begin{figure}
    \parbox[b]{0.48\textwidth}{
    \includegraphics[totalheight=6cm, bb = 90 35 690 575, clip =
    true]{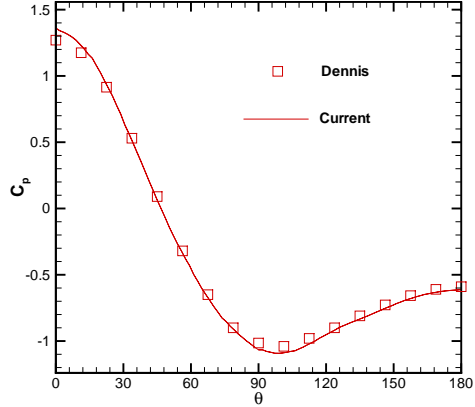}
    }
    \caption{The pressure coefficient along the circular cylinder surface.}
    \label{fig:circlePCoe}
\end{figure}

The two dimensional flow passing through a circular cylinder is considered to test Cartesian grid method for steady flows at very low Mach numbers.
The free stream Mach is about 0.08, and Reynolds numbers based on the cylinder diameter, D, is Re = 10, 20, 30.
A rectangular computational domain is used with the inlet and outlet vertical boundaries at $x = -15D$ and $x = 15D$,
and the two horizontal far field boundaries at $y = \pm15D$, respectively.
The origin coincides with the center of the cylinder.
Computations are performed using a grid of $903 \times 903$ points.
The grid size is about $0.033D$.  The curved boundaries are cut by the grid lines as shown in the figure \ref{fig:circleWakeLength} on the left. The figure on the right shows the wake length at the rear of the cylinder. The pressure coefficient is defined by $C_p = (p-p_0)/(1/2(\rho_0 U_0^2))$,
where the subscript '0' denotes the upstream or inlet flow condition. The reference data is extracted from the references \cite{Coutanceau1977,Dennis1970,Fornberg1980}. Figure \ref{fig:circlePCoe} shows the pressure coefficient along the cylinder surface. The horizontal axis represents the angle of circumference where $\theta = 0$ and $\theta = 180$ correspond to the stagnation point and the rear of cylinder respectively.
The agreement is quite satisfactory.
It should be emphasized that the flow quantities at solid boundary are quit smooth in current computation owing to the constrained weighted least square interpolation.

\subsubsection{Boundary accuracy}
\begin{figure}
    \parbox[b]{0.48\textwidth}{
    \includegraphics[totalheight=6cm, bb = 90 35 690 575, clip =
    true]{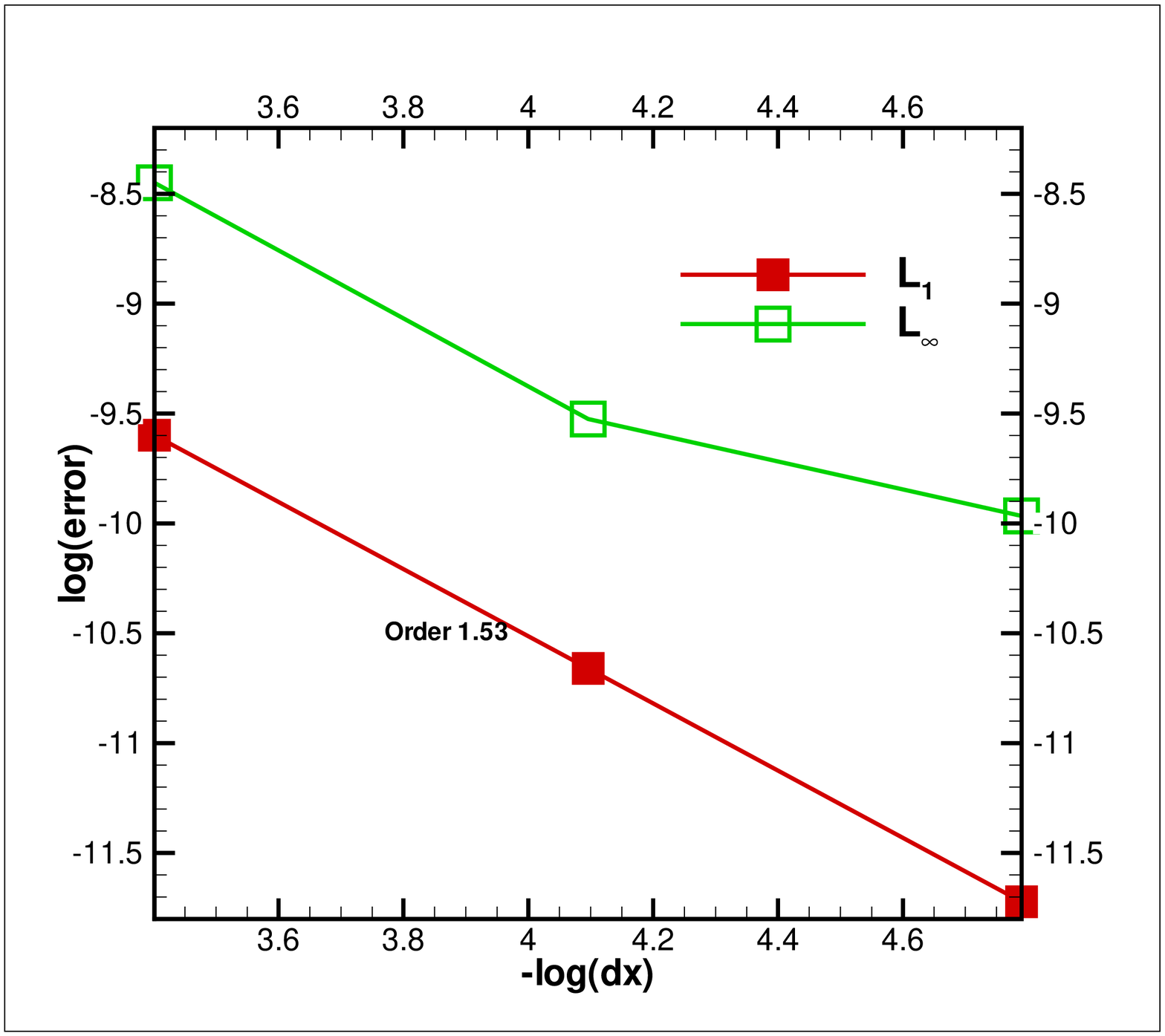}
    }
    \caption{The pressure discrepancy between the upper and lower surfaces of the cylinder vs. grid size on log scale.}
    \label{fig:accuracy} 
\end{figure}

The low Reynolds number flow passing through a circular cylinder is planar symmetric in previous example.
Therefore, the flow quantities should be exactly identical from the upper and bottom half cylinder surfaces.
However, if the cylinder is not placed symmetrically (mirror reflection relative to the x-axis), such as a small shift of the cylinder vertically,
the geometrical locations of the upper and lower half cylinder surfaces will not be symmetric.
The error introduced from this kind of discrepancy can be considered as a measure of the accuracy of the Cartesian grid method.
The $L_{1}$ error is defined as $Err_1 = \frac{1}{\pi}\int_{0}^{\pi} |P(\theta)-P(2\pi-\theta)| d\theta $; the $L_{\infty}$ error is defined as $Err_{\infty} = \max_{\theta}\{P(\theta)-P(2\pi-\theta)\}$.
We calculate the same problem on three successive refined meshes, i.e., $301\times 301$, $602\times 602$, $1204\times 1204$.
The computational domain covers the area, $[-5.004, 4.996]\times[-5.01,4.99]$, and the center of the cylinder is $(0,0)$.
Obviously, the origin and the center of cylinder are mismatched. The flow condition is the same as the setting for the Reynolds 20 case in the last example.
The errors are shown on the table \ref{tab:accuracy}, and also plotted in the figure \ref{fig:accuracy}.
The current Cartesian grid method achieves an overall second order accuracy, and has only first order accuracy at some irregular points.
This example totally excludes the influence of the flow solver.
Thus, it is a good example for the accuracy evaluation for any Cartesian grid method.
\begin{table}
\centering
\begin{tabular}{ccc}
  \hline
  grid size (d$x$)    & $L_1$ error    & $L_{\infty}$ error \\
  \hline
  \hline
  $0.03333$ & $6.781\times 10^{-5}$ & $2.140\times 10^{-4}$ \\
  $0.01664$ & $2.346\times 10^{-5}$ & $7.302\times 10^{-5}$ \\
  $0.00831$ & $8.115\times 10^{-6}$ & $4.697\times 10^{-5}$ \\
  \hline
\end{tabular}
  \caption{The pressure discrepancy between the upper and lower surfaces of the cylinder for the symmetric flow on the asymmetric distributed meshes at Re = 20.}
\label{tab:accuracy}
\end{table}

\subsubsection{Flow over a square cylinder}
\begin{figure}
    \parbox[b]{0.48\textwidth}{
    \includegraphics[totalheight=6cm, bb = 90 35 690 575, clip =
    true]{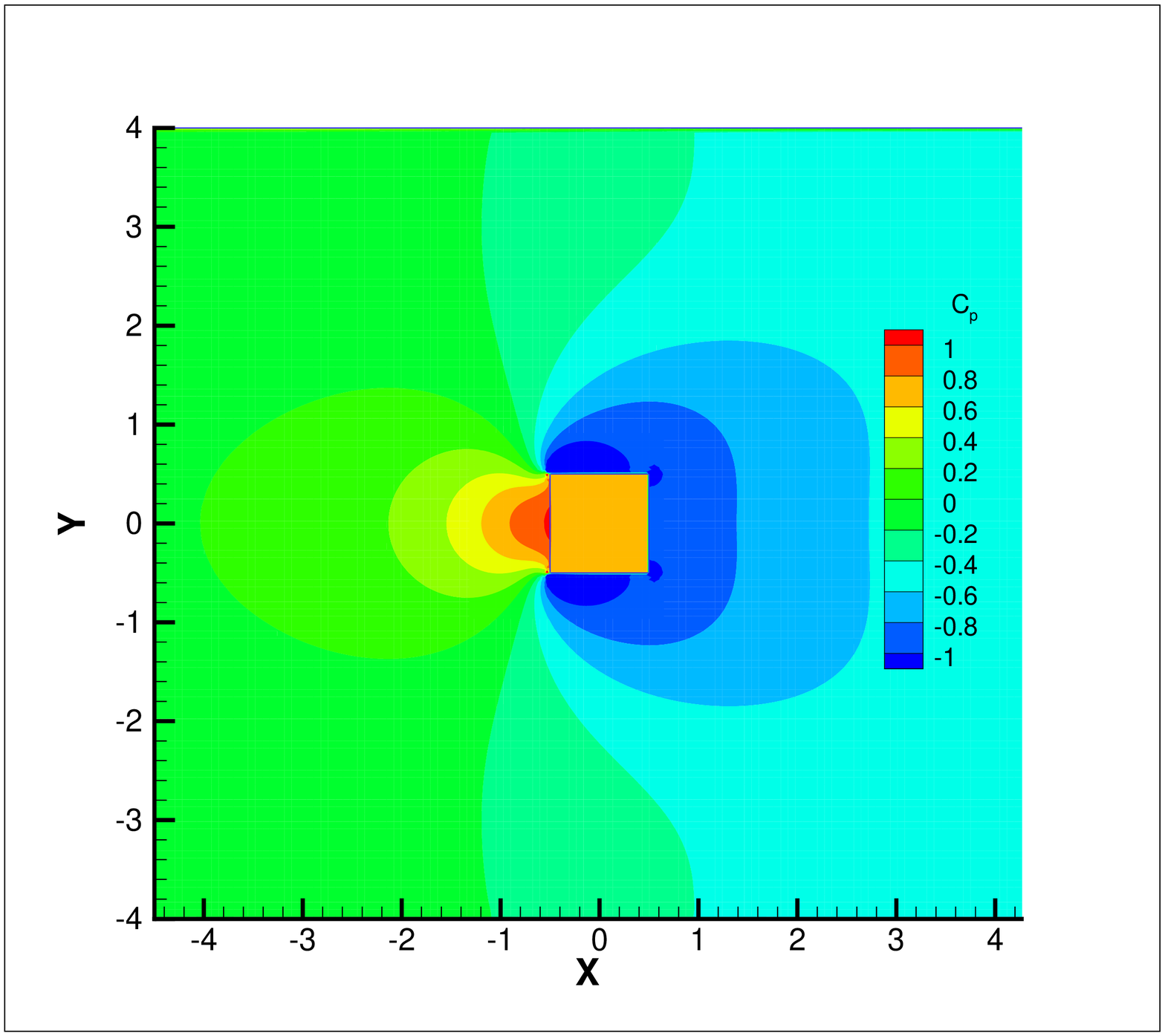}
    }
    \hfill
    \parbox[b]{0.48\textwidth}{
    \includegraphics[totalheight=6cm, bb = 90 35 690 575, clip =
    true]{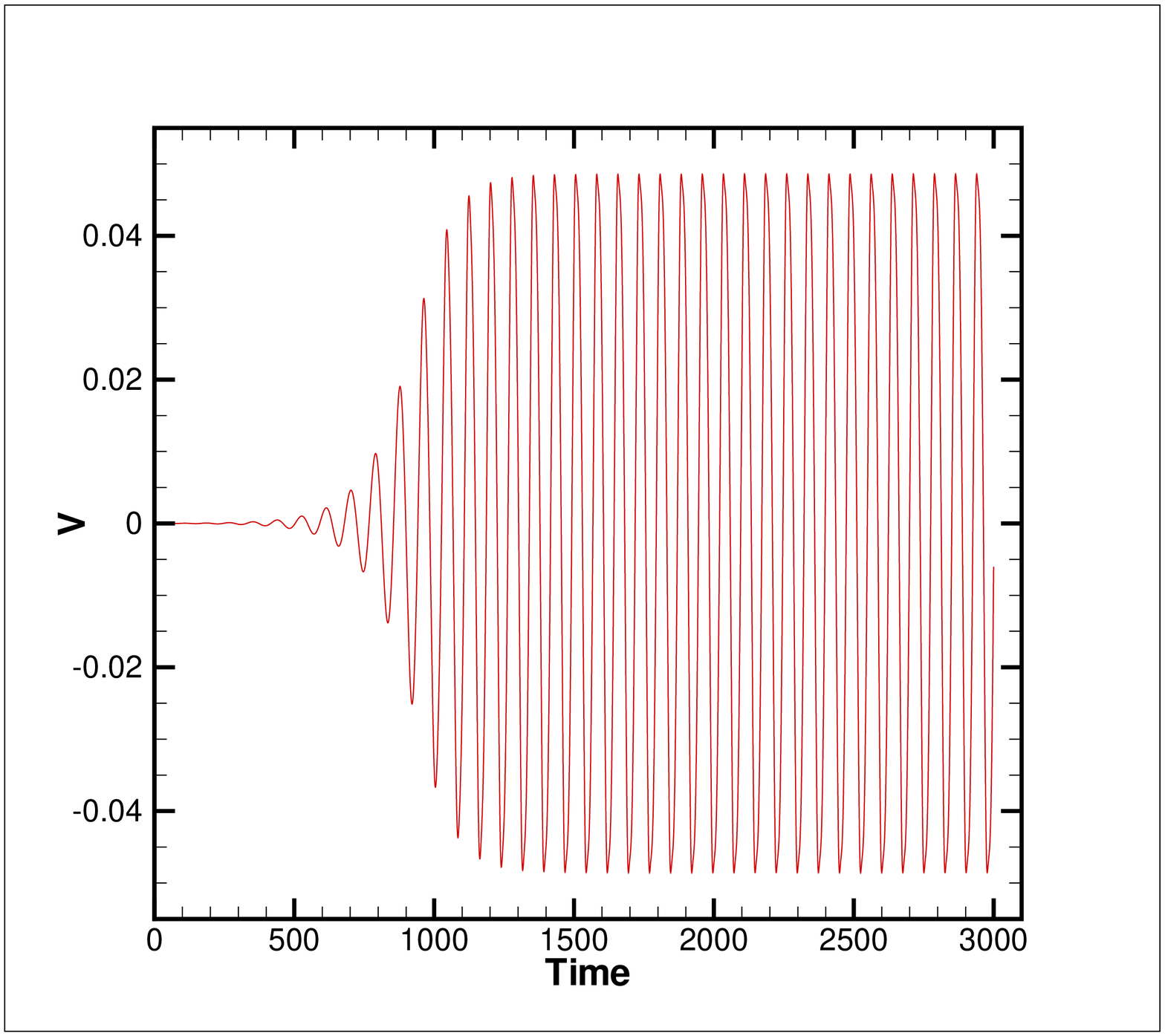}
    }
    \caption{The left figure is the pressure coefficient contour when Re = 20.
    The right figure is the vertical velocity series at the location $(3.0,0.0)$ behind the square when Re = 100. The corresponding Strouhal number is 0.137.}
    \label{fig:squareWakePattern} 
\end{figure}

\begin{figure}
    \parbox[b]{0.48\textwidth}{
    \includegraphics[totalheight=6cm, bb = 90 35 690 575, clip =
    true]{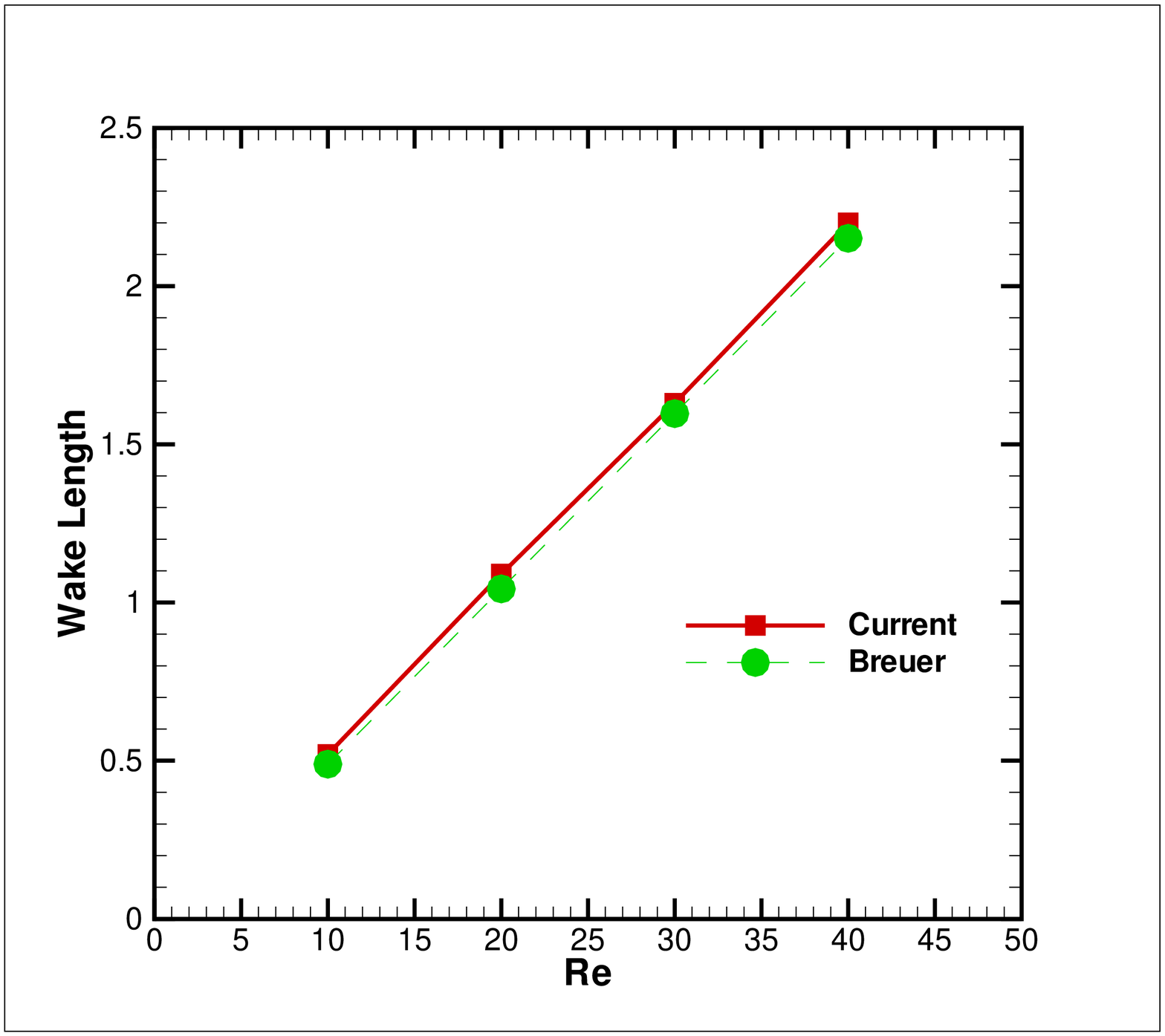}
    }
    \caption{The wake length vs. Reynolds number for the flow past a square cylinder.}
    \label{fig:squareWakeLength} 
\end{figure}

The 2D laminar flow around a square cylinder with length D mounted at the center of the channel is investigated.
The configuration is the same as the reference \cite{Breuer2000}.
The blockage ratio is fixed at B = 1/8. So the height of channel is 8D.
In order to reduce the influence of inflow and outflow boundary conditions, the length of the channel is set to be L/D = 35.
The square cylinder locates at the length of 10D from the inflow boundary. The grid is $1416\times 341$ points.
The incoming flow's velocity profile is given as a parabolic curve. The maximum velocity in the middle of channel is 0.1. The initial gas density is uniformly 1.  The inlet pressure is 1, and the pressure of outlet is given based on a typical pressure drop for tube flow.
Figure \ref{fig:squareWakePattern} shows the pressure coefficient contour at $\mbox{Re} = 20$ and the time series of the unsteady vertical velocity at location $(3.0, 0.0)$ when $\mbox{Re} = 100$. The Strouhal number derived from current method is about 0.137, which is identical to the reference data \cite{Breuer2000}. The wake length behind square is compared with reference data in figure \ref{fig:squareWakeLength}. It also coincides with the numerical simulation by Breuer\cite{Breuer2000}.

\subsection{Supersonic flows}

\subsubsection{Supersonic flow around a circular cylinder}
\begin{figure}
    \parbox[b]{0.48\textwidth}{
    \includegraphics[totalheight=6cm, bb = 90 35 690 575, clip =
    true]{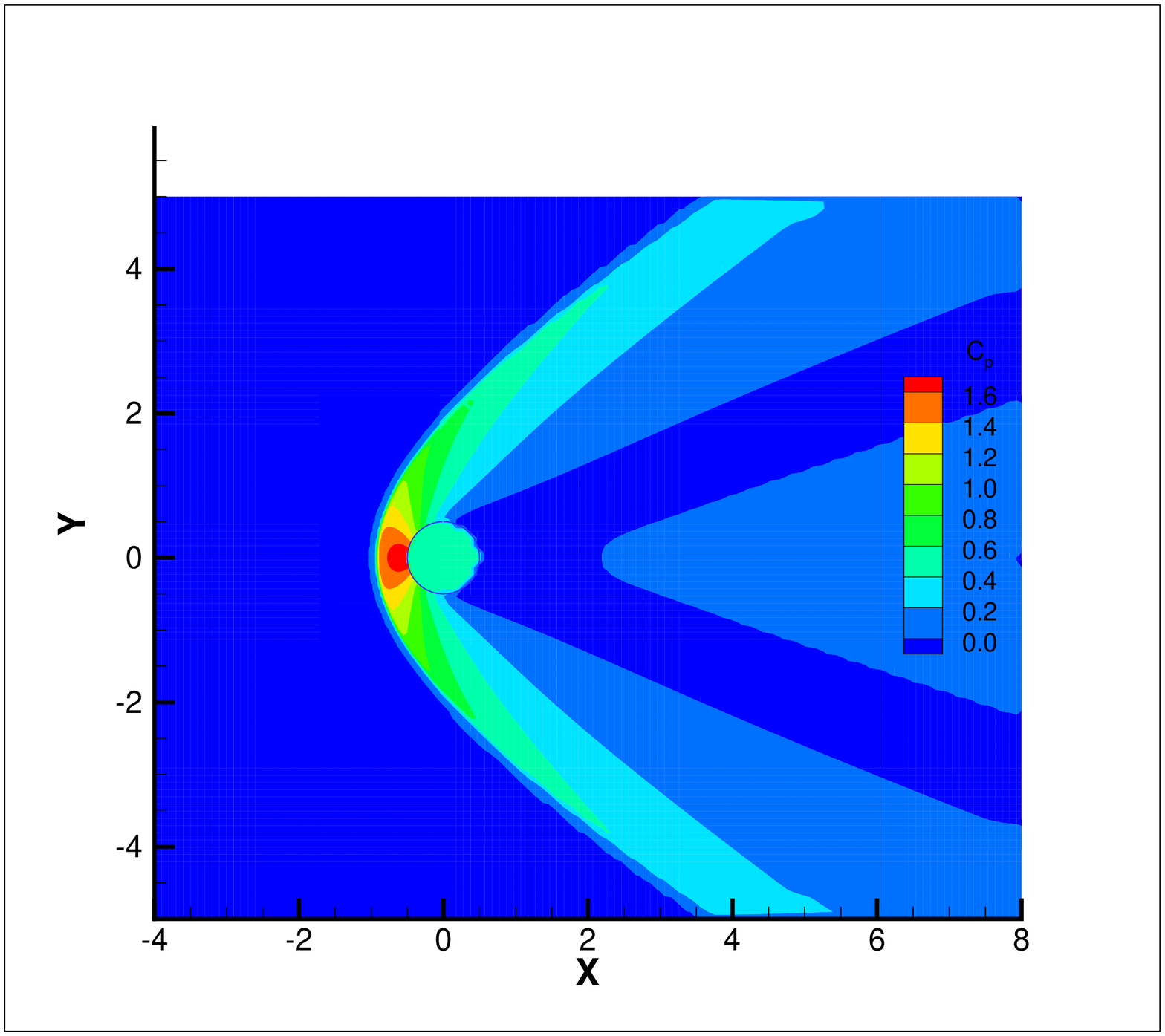}
    }
    \parbox[b]{0.48\textwidth}{
    \includegraphics[totalheight=6cm, bb = 90 35 690 575, clip =
    true]{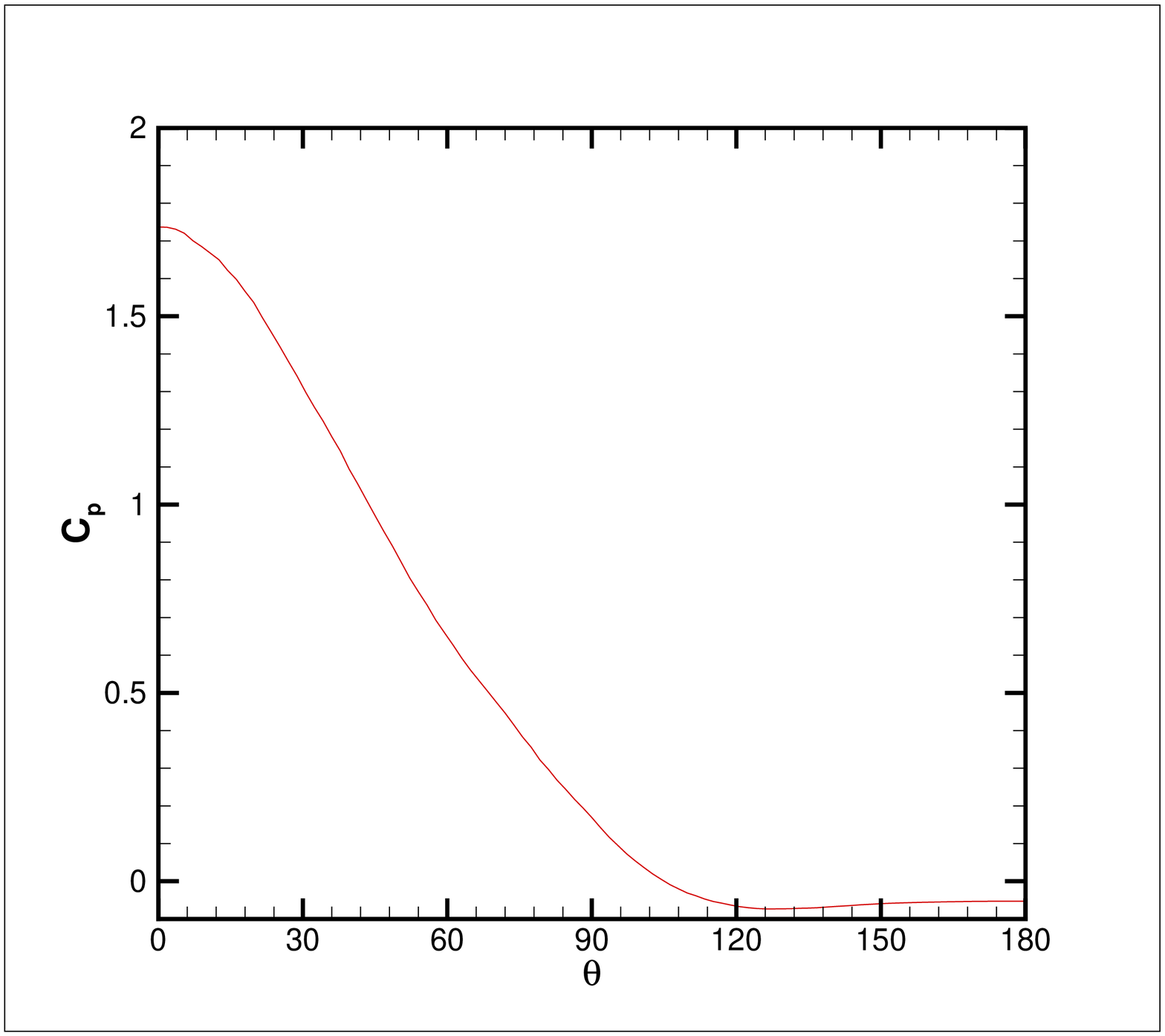}
    }
    \caption{The contour of the pressure coefficient (left) and the pressure coefficient (right) along the cylinder surface for the flow around a circular cylinder at Mach = 3.}
    \label{fig:mach3Cylinder} 
\end{figure}

\begin{figure}
    \parbox[b]{0.48\textwidth}{
    \includegraphics[width=8cm, bb = 90 35 690 575, clip =
    true]{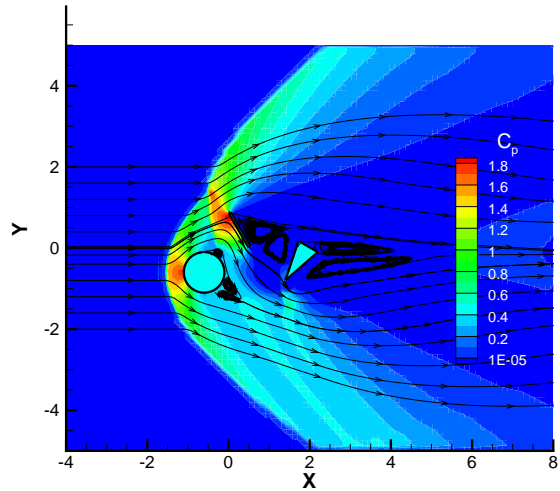}
    }
    \caption{The pressure coefficient contour and the stream lines for the flow past multiple objects when Ma = 3.}
    \label{fig:multiObjField} 
\end{figure}

Flow around a blunt body is a very important test case for supersonic flow solver.
We use this case to check the robustness of the current Cartesian grid method.
The gas initially has a velocity of Mach 3.
The Reynolds number defined by upstream condition is 1420. And the upstream pressure is 1.
The grid with $363 \times 303$ mesh points covers  the range of $[-4.0,8.0]\times[-5.0,5.0]$.
Isothermal no-slip boundary condition is employed on the cylinder surface ($T_{wall}=1$).
The scheme survives initially at the strong rarefaction wave at the rear part of the cylinder.
The figure \ref{fig:mach3Cylinder} shows the pressure coefficient contour and the pressure coefficient along the cylinder surface at the steady state.
Then we put another two different geometries, say, a plate and a triangle, into the computational field. The flow condition is identical to the previous simulation. Figure \ref{fig:multiObjField} shows the pressure coefficient contour and the stream lines. This simulation demonstrates the robustness and flexibility of current Cartesian grid method.

\subsubsection{Mach 3 forward step flow}
\begin{figure}
\centering
    \parbox[t]{0.2\textwidth}{
    \includegraphics[width=4cm, bb = 162 195 761 708, clip =
    true]{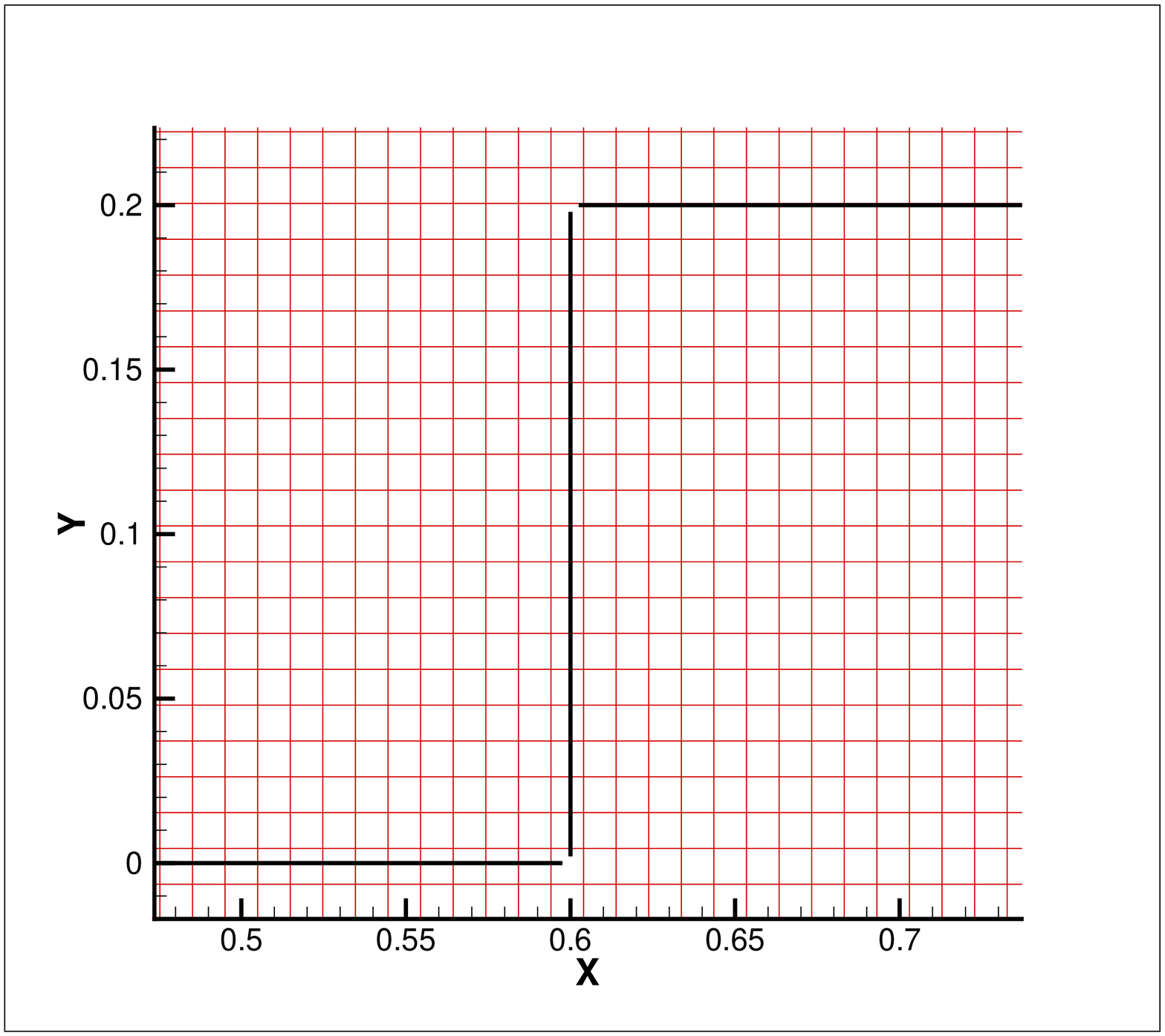}
    }
    \parbox[t]{0.7\textwidth}{
    \includegraphics[width=10cm, bb = 117 313 820 567, clip =
    true]{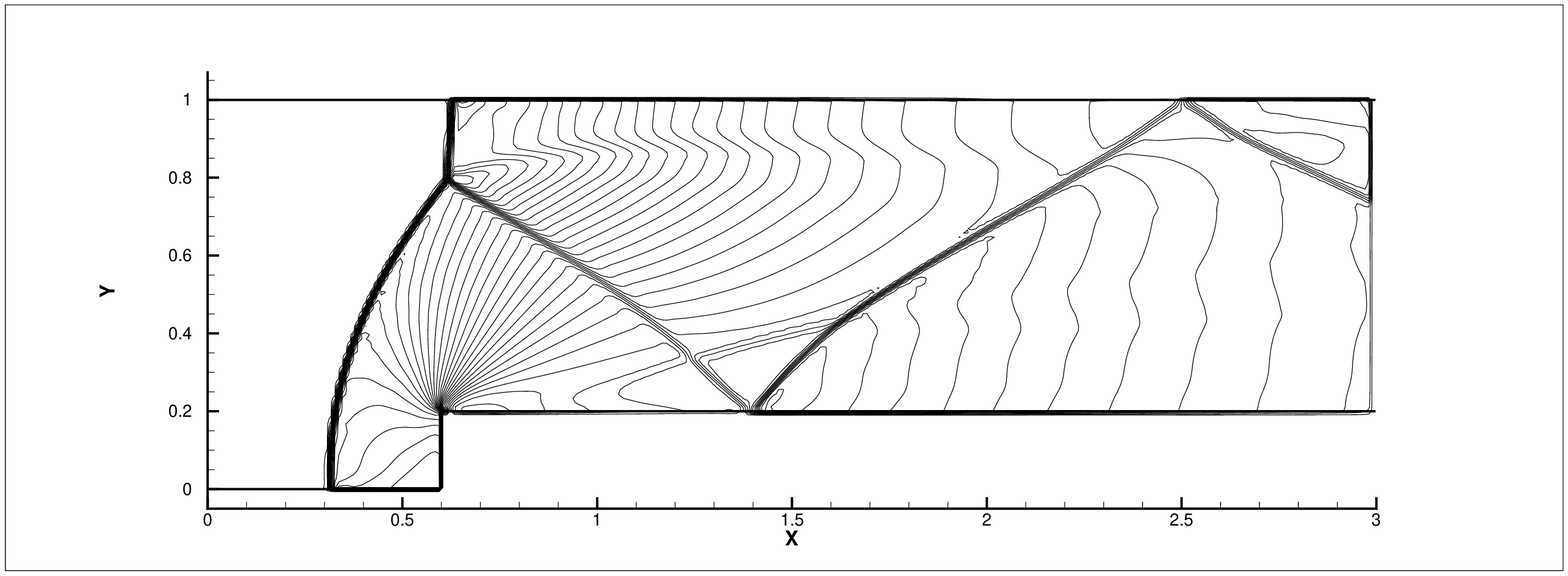}
    \includegraphics[width=10cm, bb = 117 313 820 567, clip =
    true]{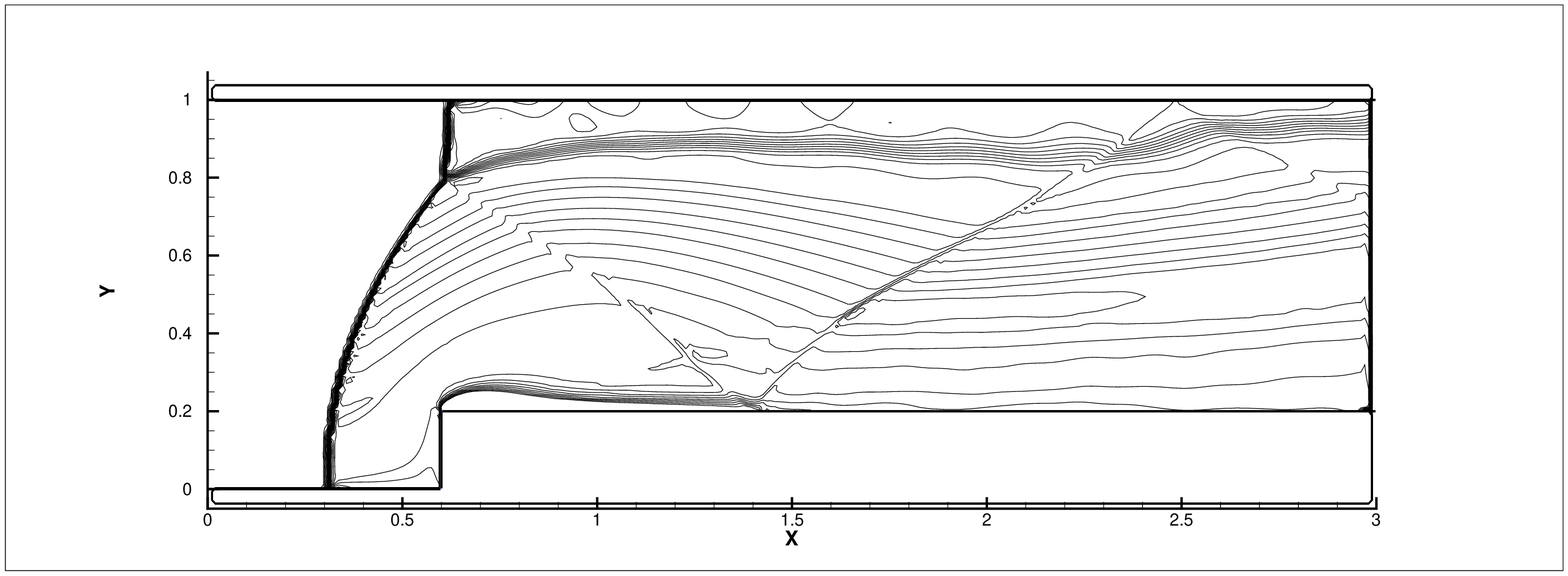}
    }
    \caption{The left figure illustrates the mesh around the step. The right top figure shows the density contour for Mach 3 step flow. The right bottom figure is the contour of $p/\rho^{\gamma}$. The grid is $304 \times 102$ covering the range of $[0.0,3]\times[-0.05,1.05]$. }
    \label{fig:forwardStep} 
\end{figure}

This two dimensional test problem has been studied by Chiang \emph{et al.} \cite{Chiang1992} using Cartesian grid method. However, in their figures, the grid lines are superposed on the boundary. In our simulation, the computational
domain is $[0.0, 3.0] \times [-0.05, 1.05]$. Two horizontal boundaries are placed at $y = 0.0$ and $y = 1.0$ respectively.
A step with height 0.2 is located at x = 0.6. The upstream velocity is $(U, V) = (3, 0)$. The gas in the computational domain is uniform at the beginning, whose density is 1.4, pressure is 1.0, and velocity is 3. The Euler slip boundary condition is imposed
at all solid boundaries. The horizontal direction is discretized by 304 points. Then we change the number of the vertical discrete points. The length of Mach stem at the top surface varies when the vertical discrete point number changes from 100 to 110. This is because that the minimum distance to the top boundary changes dramatically when in the vertical discrete point number changes. And if double the mesh points in each direction, the numerical solution becomes insensitive to the small changes of the vertical discrete point number.
Finally, the solution indeed converges to the reference solution\cite{Woodward1984} when the mesh is fine enough even if the shock exists near the solid boundary.
The density distribution in the wind tunnel and the entropy ($p/\rho^{\gamma}$) at time $t=4$ are shown in figure \ref{fig:forwardStep}. No special treatment is adopted at the corner of the step.

\subsubsection{Double Mach reflection}
\begin{figure}
\centering
    \parbox[b]{0.8\textwidth}{
    \centering
    \includegraphics[width=0.4\textwidth, bb = 90 35 690 575, clip =
    true]{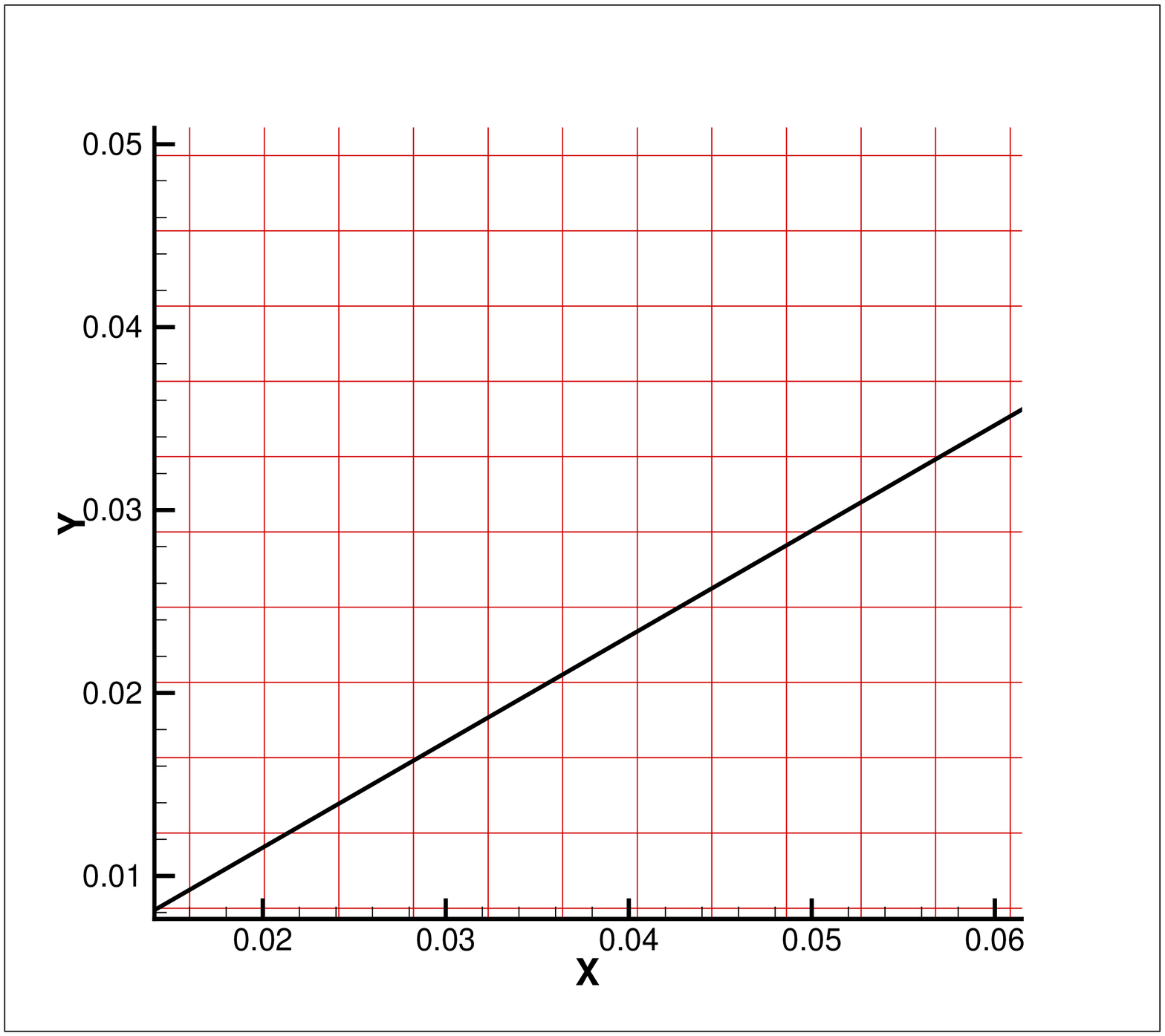}
    }
    \parbox[b]{0.48\textwidth}{
    \includegraphics[width=0.48\textwidth, bb = 97 178 610 560, clip =
    true]{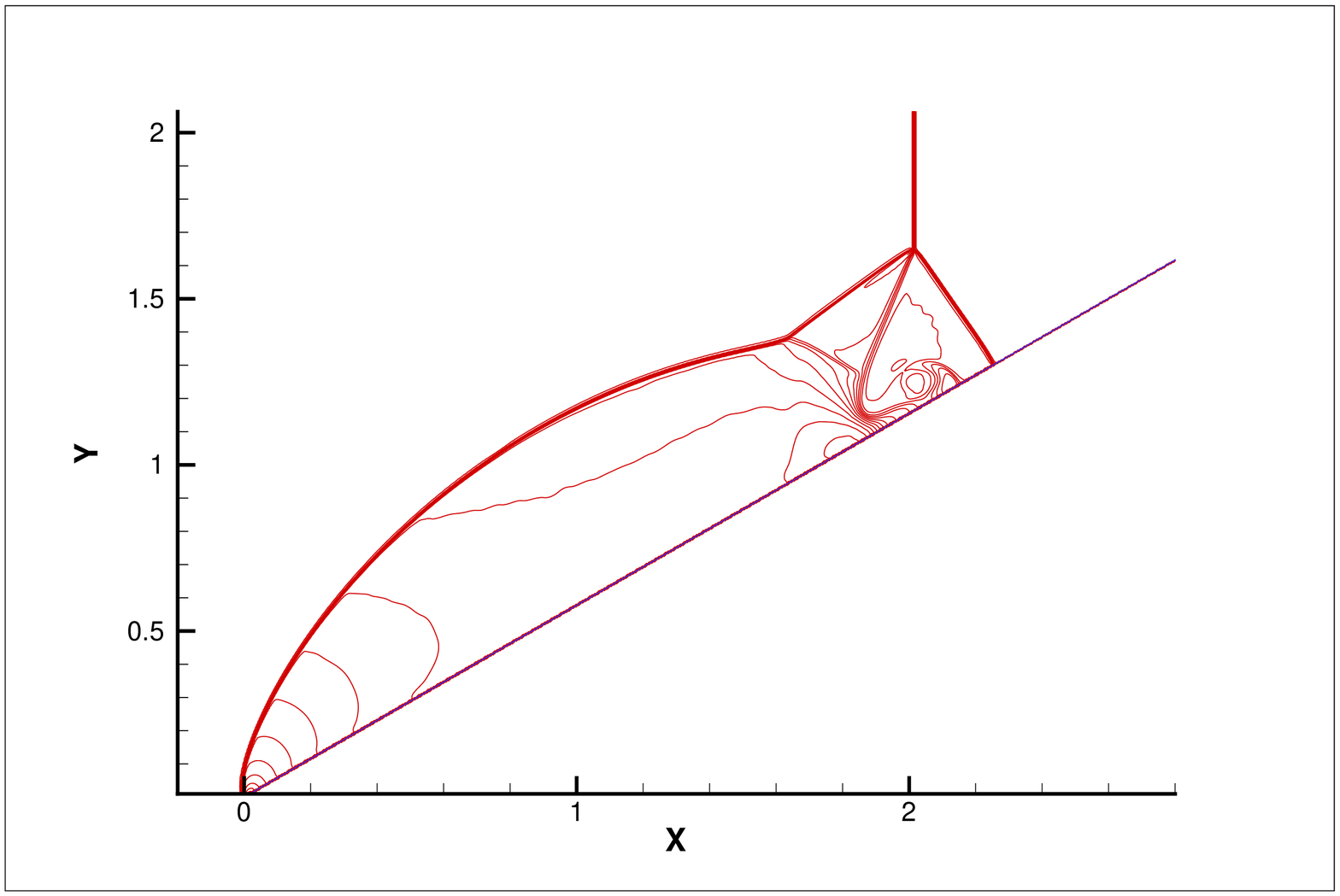}
    }
    \parbox[b]{0.48\textwidth}{
    \includegraphics[width=0.48\textwidth, bb = 97 178 610 560, clip =
    true]{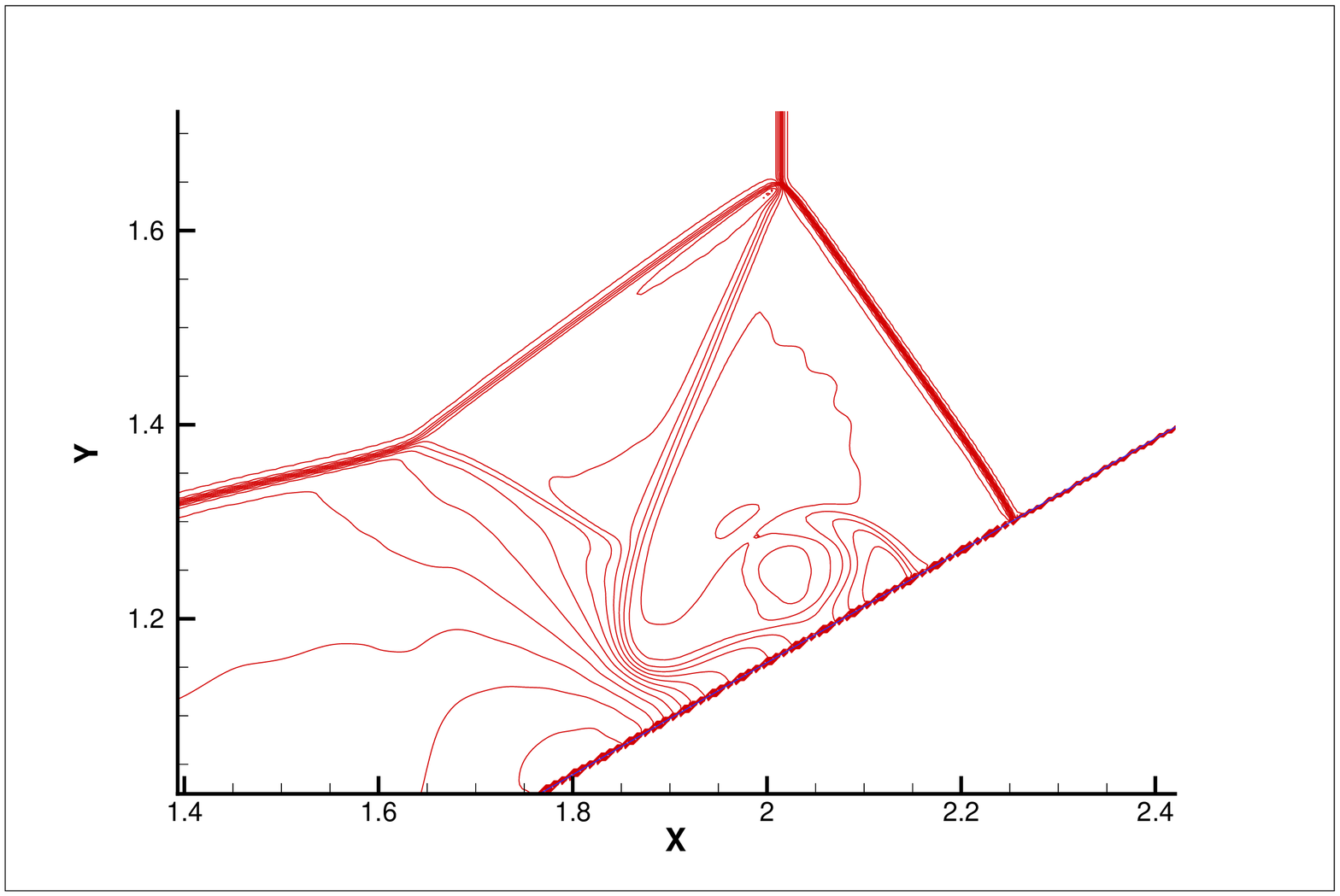}
    }
    \caption{The mesh is $900 \times 730$ covering the range of $[0.0,3]\times[-0.2,3.464]$. The bottom left figure shows a portion of the mesh. The bottom right figure shows the density contour of double Mach reflection.}
    \label{fig:doubleMach} 
\end{figure}

A wedge is placed at the origin with 30 degree angle to $x$ direction. The computational domain is $[-0.2,3.464]\times[0.0,3.0]$. The gas at $x>0$, is at rest with density 1.4, and pressure 1. The flow condition at $x < 0$ is as follows, $\rho = 8, U = 8.25, p = 116.5$. A mesh of $900 \times 730$ points is used. The Euler slip boundary condition is imposed at the solid boundaries. The figure \ref{fig:doubleMach} displays the density contours. This case has already been studied by Cartesian grid methods \cite{Forrer1998,Pember1995}. In Forrer's work \cite{Forrer1998},
the whole structure moves slower than benchmark results. Our simulation results coincide with the results obtained by Pember \cite{Pember1995} and  the results from the calculation of a conformal mesh\cite{Qiu2003}.

\subsubsection{Viscous shock tube}
\begin{figure}
\centering
    \parbox[b]{0.2\textwidth}{
    \includegraphics[width=0.2\textwidth, bb = 101 44 705 550, clip =
    true]{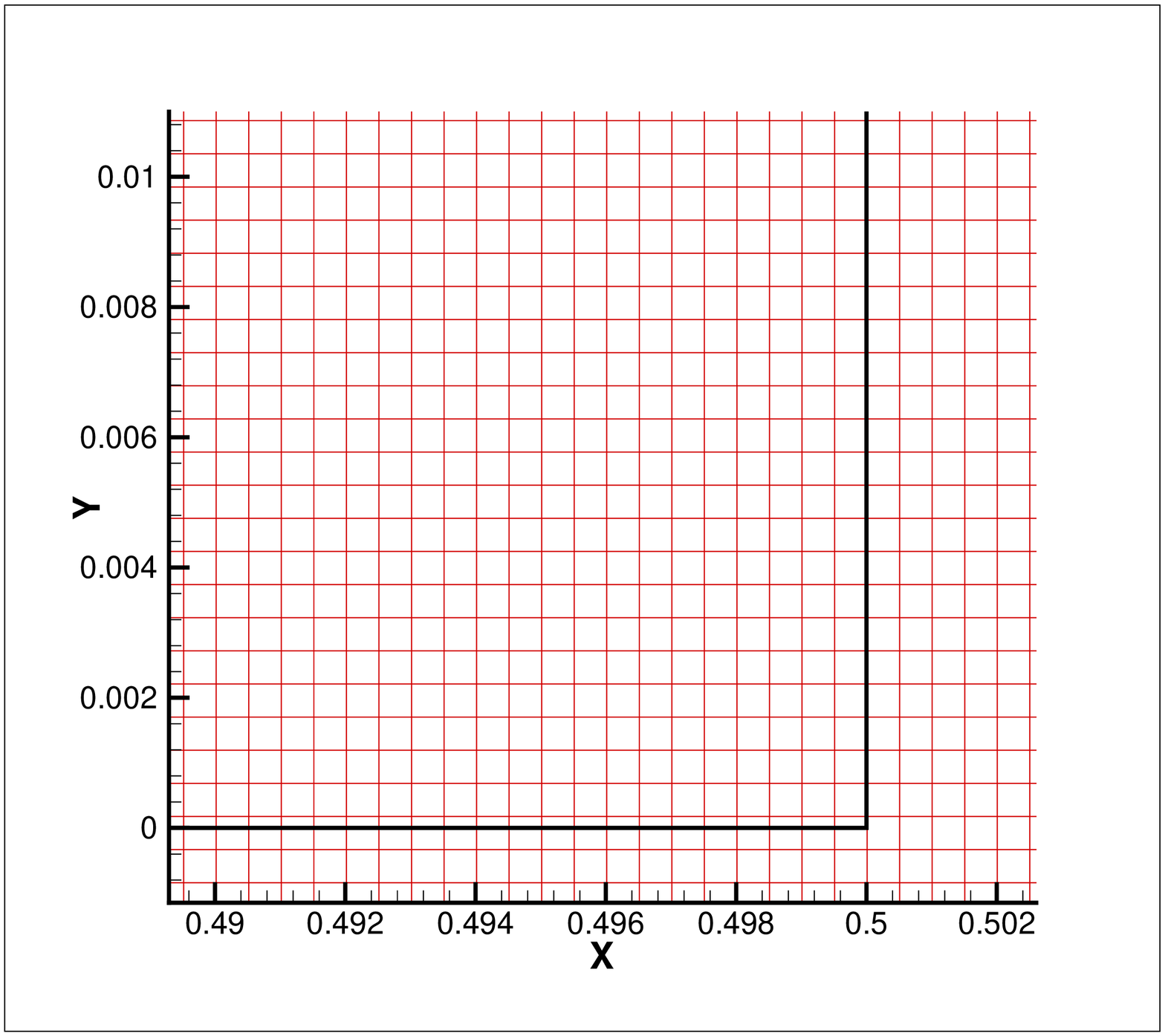}
    }
    \parbox[b]{0.7\textwidth}{
    \includegraphics[width=0.7\textwidth, bb = 104 314 660 560, clip =
    true]{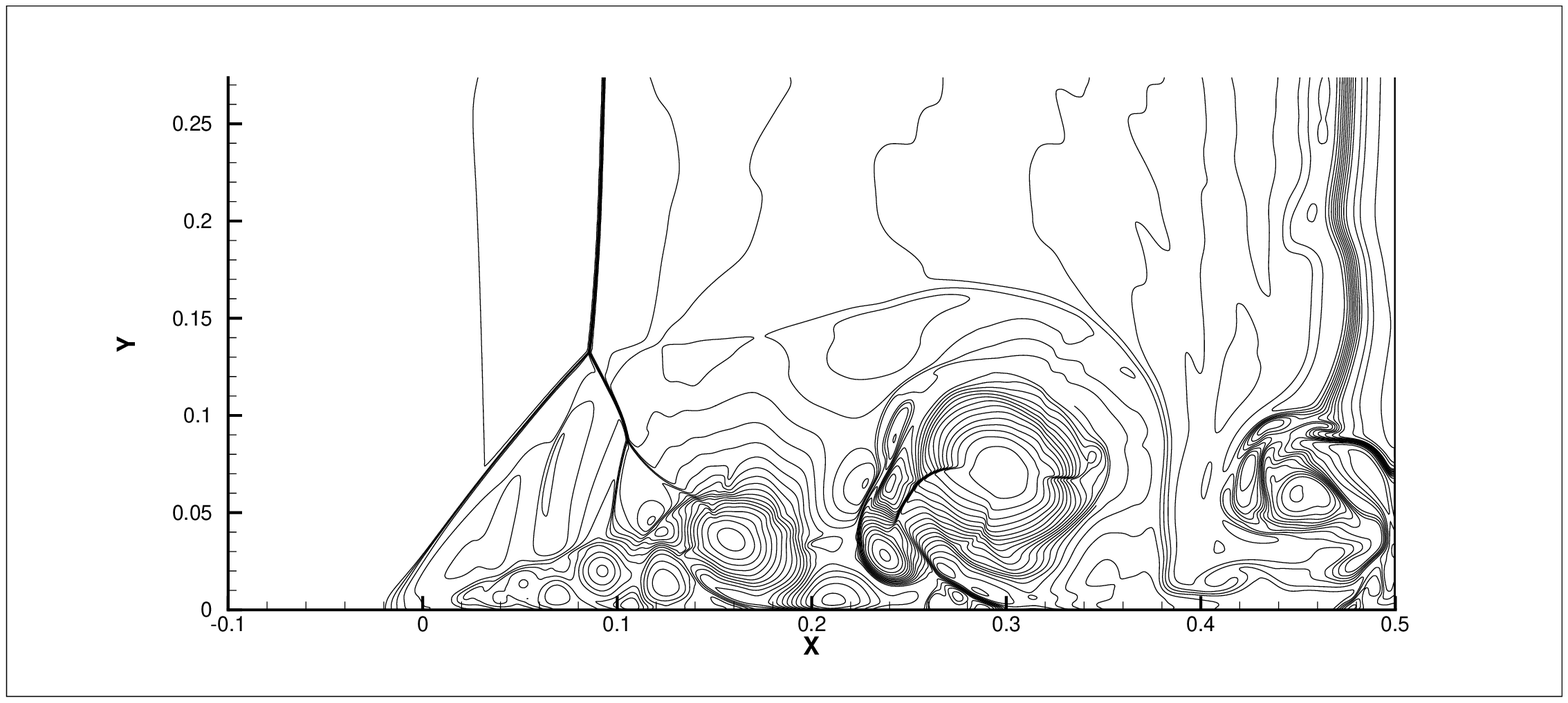}
    }
    \caption{The left figure shows the mesh around the right bottom corner of the computational domain. The right figure shows the density contour of viscous shock tube. The grid is $2043 \times 1023$ covering the range of $[-0.51,0.51]\times[-0.01,0.51]$ }
    \label{fig:visShockTube} 
\end{figure}

This is a viscous problem introduced by Daru and Tenaud \cite{Daru2001,Daru2009}. Sj{\"o}green and
Yee \cite{Sjogreen2003} and many other researchers also studied this problem. The gas with different thermal condition is initially at rest on both sides of $x=0.5$, and separated by a membrane. Then, the membrane is removed and wave interaction occurs. The compressible Navier-Stokes equations
with adiabatic no-slip boundary conditions are imposed. The solution develops complex two dimension
shock/shear/boundary-layer interactions, which depend on the Reynolds number.
The dimensionless initial states is given as follows,
$$\rho_l = 120, p_l = 120/\gamma, \rho_r = 1.2, p_r = 1.2/\gamma ,$$
where subscripts "l" and "r" denotes the left side and right side of $x = 0.5$ respectively. The Prandtl number
is 0.73. The case of Reynolds number 1000 to time $t = 1.0$ is simulated. The computational
domain is $[-0.51, 0.51] \times [-0.01, 0.51]$. A box of $[-0.5, 0.5] \times [0.0, 0.5]$ embraces the fluid domain.
$2043\times 1023$ grid points are used in this simulation.
The bottom and two lateral boundaries are adiabatic with non-slip condition. The top boundary is reflection boundary (Euler slip boundary).
Generally speaking, the structures of this problem (figure \ref{fig:visShockTube}) are quit similar to that derived in references \cite{Daru2009,Sjogreen2003}.
This simulation demonstrates that current Cartesian grid method is capable of simulating complex supersonic viscous flow.

\section{Conclusion}
In this paper, we present a Cartesian grid method for complex immersed boundary problems, and propose a simplified gas kinetic scheme for subsonic and supersonic flow simulation.
A constrained weighted least square method is employed to update the physical quantities at the interpolation points.
Different boundary conditions, including isothermal, adiabatic, and Euler slip conditions, are presented by different interpolation strategies.
The new method is capable of simulating inviscid and viscous, compressible and near incompressible flow problems.
The numerical results demonstrate that the new method converges to the correct physical solutions for both subsonic and supersonic flows as the mesh refines.
The current scheme is robust under various flow condition from low Mach number to high Mach number flows, from inviscid to viscous ones.
All these test cases are calculated by the same Cartesian grid method.
The interpolation procedure proposed in this study provides a smooth distribution of physical quantities at solid boundary and can tackle with arbitrary geometry.
The current methodology can be further extended to the other flow systems, such as to the numerical schemes \cite{ugks1_1,chen_adap,ugks3,Xubook,Li2004,Li2009} for both continuum and rarefied flow computations.

\vspace{3ex} {\textbf{Acknowledgments}} \vspace{1ex}
This work was supported by Hong Kong Research Grant Council (621011,620813,16211014), grants IRS15SC29 and SBI14SC11 at HKUST, National Nature Science Foundation of China under Grant Nos.91130018,11325212 and National Key Basic Research and Development Program (2014CB744100).

\bibliographystyle{elsarticle-num}
\bibliography{CartesianMesh}


\end{document}